\documentclass[iop,twocolumn,apj]{emulateapj}

\usepackage{apjfonts}
\usepackage{natbib,graphicx,epsfig,rotating,color,verbatim,amsmath,booktabs}

\usepackage{bm}
\usepackage{hyperref}

\newcommand{\beq}{\begin{equation}}
\newcommand{\eeq}{\end{equation}}
\newcommand{\lsi}{\,\raisebox{-0.13cm}{$\stackrel{\textstyle<}
{\textstyle\sim}$}\,}

\newcommand{\be}{\begin{equation}}
\newcommand{\ee}{\end{equation}}
\newcommand{\bea}{\begin{eqnarray}}
\newcommand{\eea}{\end{eqnarray}}

\newcommand{\bw}{\begin{widetext}}
\newcommand{\ew}{\end{widetext}}

\def\<{\langle}
\def\>{\rangle}

\def\alt{\raise0.3ex\hbox{$\;<$\kern-0.75em\raise-1.1ex\hbox{$\sim\;$}}}
\def\agt{\raise0.3ex\hbox{$\;>$\kern-0.75em\raise-1.1ex\hbox{$\sim\;$}}}

\def\deg{\hbox{$^\circ$}}

\begin{document}
\title{Constraints on the Galactic Halo Dark Matter from Fermi-LAT Diffuse Measurements}

%\author{The Fermi-LAT Collaboration}
%\affiliation{{\it ...}}
%%%%%%%%%%%%%%%%%%%%%%%%%%%%%%%%%%%%%%%%%%
\author{
M.~Ackermann\altaffilmark{1}, 
M.~Ajello\altaffilmark{2,3}, 
W.~B.~Atwood\altaffilmark{4}, 
L.~Baldini\altaffilmark{5}, 
G.~Barbiellini\altaffilmark{6,7}, 
D.~Bastieri\altaffilmark{8,9}, 
K.~Bechtol\altaffilmark{2}, 
R.~Bellazzini\altaffilmark{10}, 
R.~D.~Blandford\altaffilmark{2}, 
E.~D.~Bloom\altaffilmark{2}
% ...
}
\altaffiltext{1}{Deutsches Elektronen Synchrotron DESY, D-15738 Zeuthen, Germany}
\altaffiltext{2}{W. W. Hansen Experimental Physics Laboratory, Kavli Institute for Particle Astrophysics and Cosmology, Department of Physics and SLAC National Accelerator Laboratory, Stanford University, Stanford, CA 94305, USA}
\altaffiltext{3}{Department of Astronomy, University of California, Berkeley, CA 94720-3411, USA}
\altaffiltext{4}{Santa Cruz Institute for Particle Physics, Department of Physics and Department of Astronomy and Astrophysics, University of California at Santa Cruz, Santa Cruz, CA 95064, USA}
\altaffiltext{5}{Universit\`a  di Pisa and Istituto Nazionale di Fisica Nucleare, Sezione di Pisa I-56127 Pisa, Italy}
\altaffiltext{6}{Istituto Nazionale di Fisica Nucleare, Sezione di Trieste, I-34127 Trieste, Italy}
\altaffiltext{7}{Dipartimento di Fisica, Universit\`a di Trieste, I-34127 Trieste, Italy}
\altaffiltext{8}{Istituto Nazionale di Fisica Nucleare, Sezione di Padova, I-35131 Padova, Italy}
\altaffiltext{9}{Dipartimento di Fisica e Astronomia "G. Galilei", Universit\`a di Padova, I-35131 Padova, Italy}
\altaffiltext{10}{Istituto Nazionale di Fisica Nucleare, Sezione di Pisa, I-56127 Pisa, Italy}
% ...
%\section{}
%%%%%%%%%%%%%%%%%%%%%%%%%%%%%%%%%%%%%%%%%%%

%\linenumbers
\begin{abstract}
We have performed an analysis of the diffuse gamma-ray emission with the {\it Fermi} Large Area Telescope in the Milky Way Halo  region 
searching for a signal from  dark matter annihilation or decay.   In the absence of a robust dark matter signal, constraints are presented.  We consider both gamma rays produced directly in the dark matter annihilation/decay and produced by inverse Compton scattering of the $e^+$/$e^-$ produced in the annihilation/decay. Conservative limits are derived requiring that the dark matter signal does not exceed  the observed diffuse gamma-ray emission. A second set of more stringent limits is derived based on modeling the foreground astrophysical diffuse emission using the \texttt{GALPROP}  code.   
Uncertainties in the height of the diffusive cosmic-ray halo, the distribution of the cosmic-ray sources in the Galaxy, the index of the injection cosmic-ray electron spectrum and the column density of the interstellar gas are taken into account using a profile likelihood formalism, while the parameters governing the cosmic-ray propagation have been derived from fits to local cosmic-ray data. 
The resulting limits impact the range of particle masses over which dark matter thermal production in the early Universe is possible, and challenge the interpretation of the PAMELA/{\it Fermi}-LAT cosmic ray anomalies as annihilation of dark matter. 
\end{abstract}

\vspace{4pt}

\maketitle

\section{Introduction}\label{introduction}

The nature of Dark Matter (DM) and its properties are still unknown, despite being one of the most widely investigated topics in contemporary fundamental physics. However, current and near future experiments are probing more and more of the parameter space predicted for the most popular type of DM candidates, Weakly Interacting Massive Particles (WIMPs) (for a review see \citealp{2000RPPh...63..793B,2004PhR...405..279B,2010Natur.468..389B}). In particular, high-energy gamma-ray astronomy can be used to search for signatures of DM in the Milky Way and beyond. Gamma-rays are products of hadronization and radiative loss processes, and are therefore unavoidably emitted in  annihilation and decay {of WIMPs}. 
The propagation of gamma rays is mostly unaffected by the interstellar medium and Galactic magnetic fields, and therefore the data retain information on the morphology of the emission region (unlike, e.g., charged cosmic rays). The Large Area Telescope (LAT), onboard the {\it Fermi} gamma-ray observatory \citep{2009ApJ...697.1071A}, is now providing  unprecedented high quality gamma-ray data.

We focus here on DM signatures in the diffuse gamma-ray emission as measured by   {\it Fermi}-LAT. About $90\%$ of the LAT photons are of diffuse origin. The {\em Galactic component} encodes information on the propagation and origin of cosmic rays, distribution of cosmic-ray sources, the interstellar medium, magnetic and radiation fields in our Galaxy and unresolved point sources, while its {\em extragalactic component} provides a signature of energetic phenomena on cosmological scales. {\em Both} components are expected to include a contribution from DM annihilation/decay: the  Galactic signal arises from the smooth DM halo around the Galactic Center and Galactic substructures, while the extragalactic one arises from the signal of DM annihilation processes throughout the Universe integrated over all redshifts. The extragalactic component is analyzed elsewhere \citep{2010JCAP...04..014A}. Here we focus on searching for  
a potential signal from DM annihilation/decay in the halo of the Milky Way (for previous work related to this topic see  \citealp{2009PhRvD..80b3007Z,2010JCAP...03..014P,2010NuPhB.840..284C,2011PhRvD..84b3013M,2011PhRvD..83l3516B,2010PhRvD..82b3518L,2011PhRvD..84b2004A}).

Due to the bright sources present in the Galactic Center and the bright diffuse emission along the plane, it has been argued, e.g. in \cite{2008APh....29..380S}, that the region of the inner Galaxy, extending $10\deg$--$20\deg$ away from the Galactic Plane, is promising in terms of the signal-to-background ratio $S/N$. 
As an additional advantage, the  constraints on the DM signal in that region become less sensitive to the unknown profile of the DM halo. In particular the $S/N$ of cored profiles is only a factor $\sim 2$ weaker than for the Navarro-Frenk-White (NFW) profile  \citep{1996ApJ...462..563N} in this region. This should be compared to an order of magnitude of uncertainty when one considers the Galactic Center region. The increase in $S/N$ ratio away from the plane is further emphasized for DM models in which DM annihilations result in a significant fraction of leptons in the final state. These leptons propagate in the Galaxy and produce high-energy gamma rays mainly through inverse Compton  scattering on the interstellar radiation field. By diffusing away from the Galactic Center region, electrons produce an extended gamma-ray signal which further enhances the S/N at higher Galactic latitudes \citep{2009ApJ...699L..59B}.

Consequently, we investigate here a large region of interest covering the central part of the Galactic Halo, while masking out the Galactic Plane. We test the LAT data for a contribution from the DM annihilation/decay signal by a fit of the spectral and spatial distributions of the expected photons in the region of interest to the LAT data. In doing so, we take into account the most up-to-date modeling of the diffuse signal of astrophysical origin \citep{2012arXiv1202.4039T}, adapting it to the problem in question.
This paper is organized in the following way: 
 {In} Sec.  \ref{diffusemodeling} {we} describe our modeling of the diffuse gamma-ray emission and the way we parameterize it. Sec. \ref{outline} outlines our general approach to fitting for DM signals in the presence of uncertainties in the astrophysical foregrounds.  
{In} Sec. \ref{maps} {we} describe the DM and gas maps used in this work {while in} Sec. \ref{sec:ROI} we define our dataset and region of interest. {In} Sec \ref{nobkg} {we derive} DM limits without modeling of the background while Sec. \ref{main} contains the details of the fit procedure for the limits which include modeling of the background. {The results are presented in} Sec. \ref{results} while  Sec. \ref{discussion} contains further discussions on the background model uncertainties. In Sec. \ref{summary} we summarize and conclude.

\begin{figure*}[t]
\begin{center}$
\begin{array}{cc}
\includegraphics[width=0.45\textwidth]{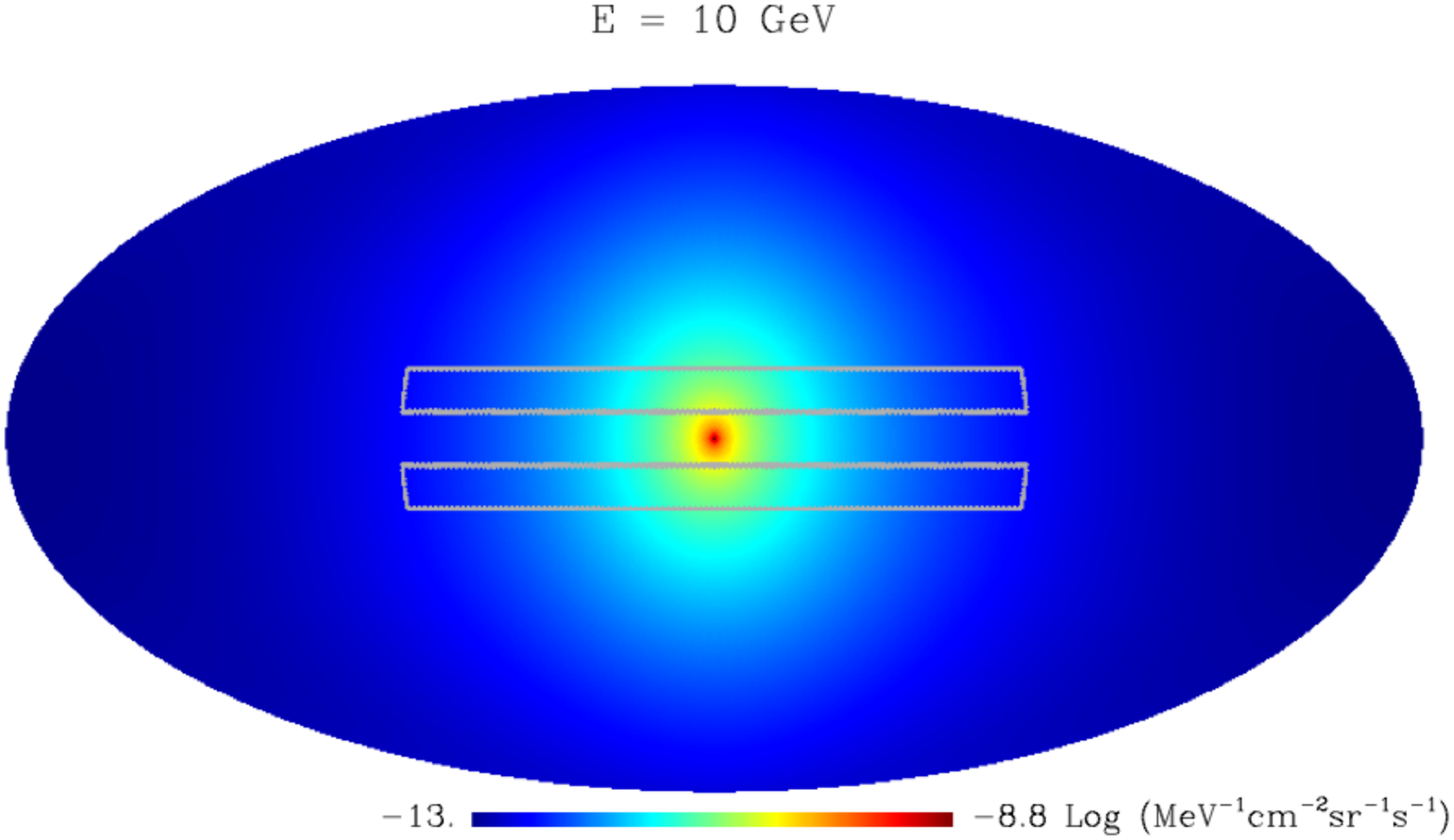}&
\includegraphics[width=0.42\textwidth]{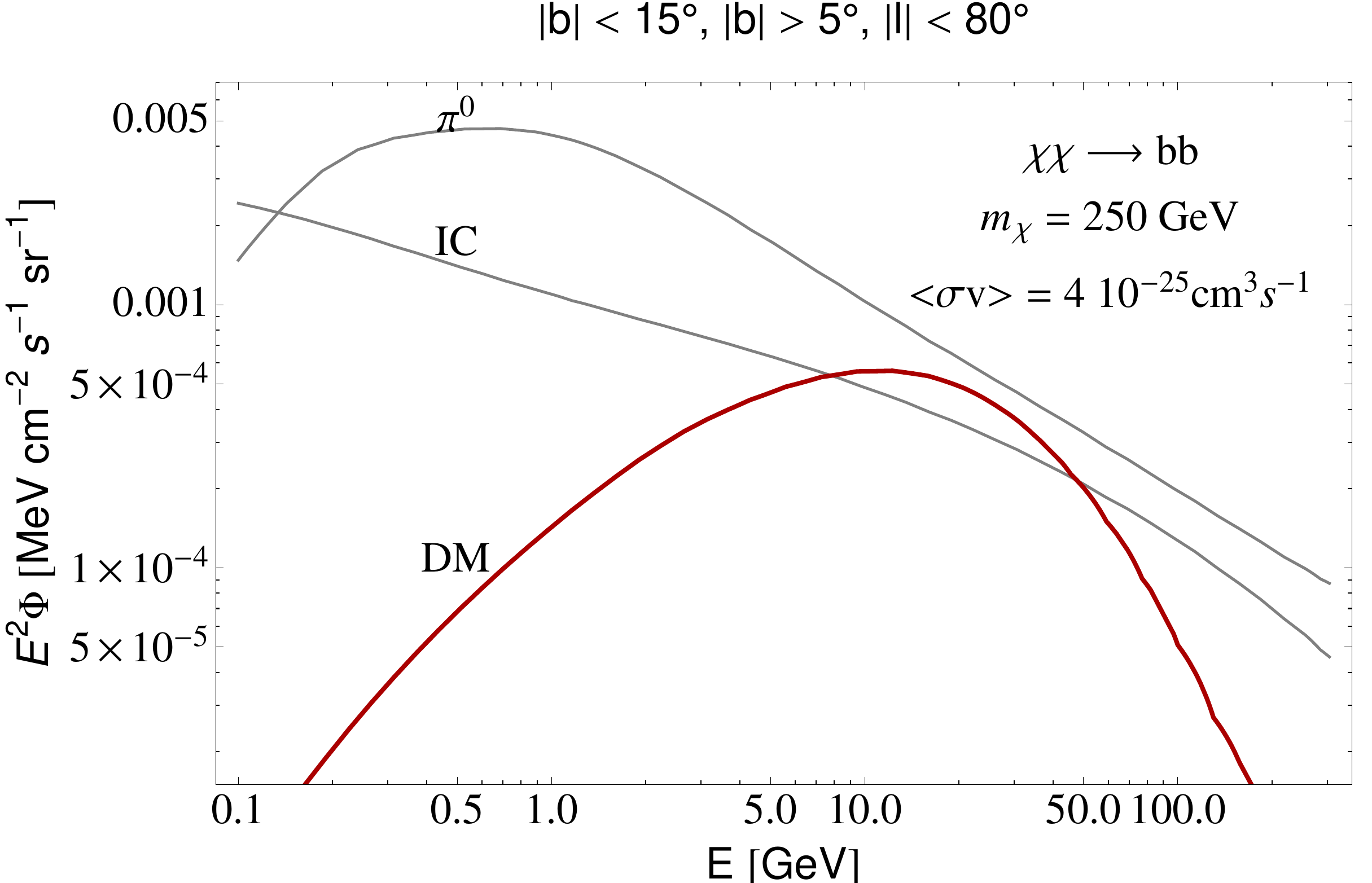} \\
\includegraphics[width=0.45\textwidth]{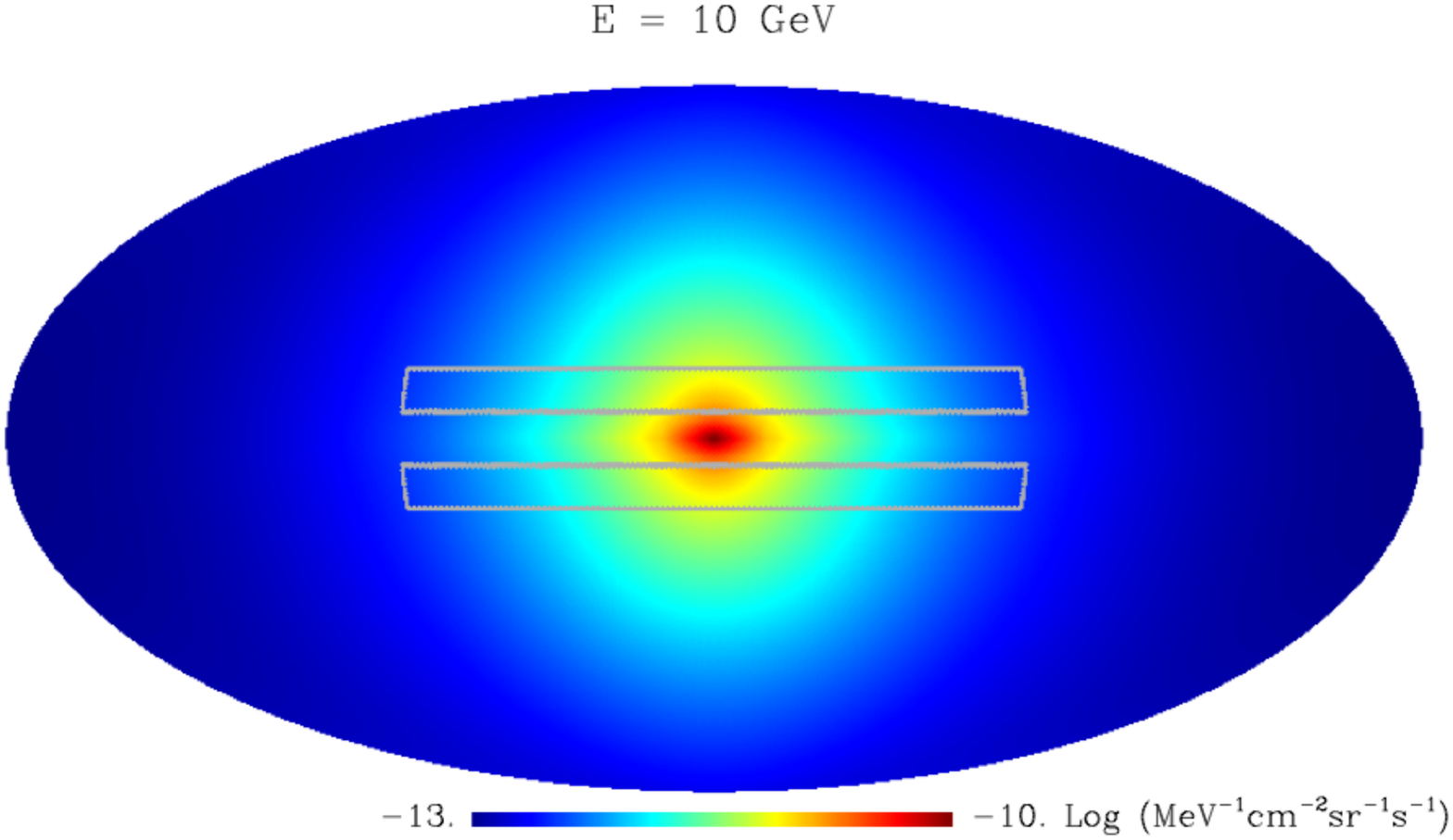}&
\includegraphics[width=0.42\textwidth]{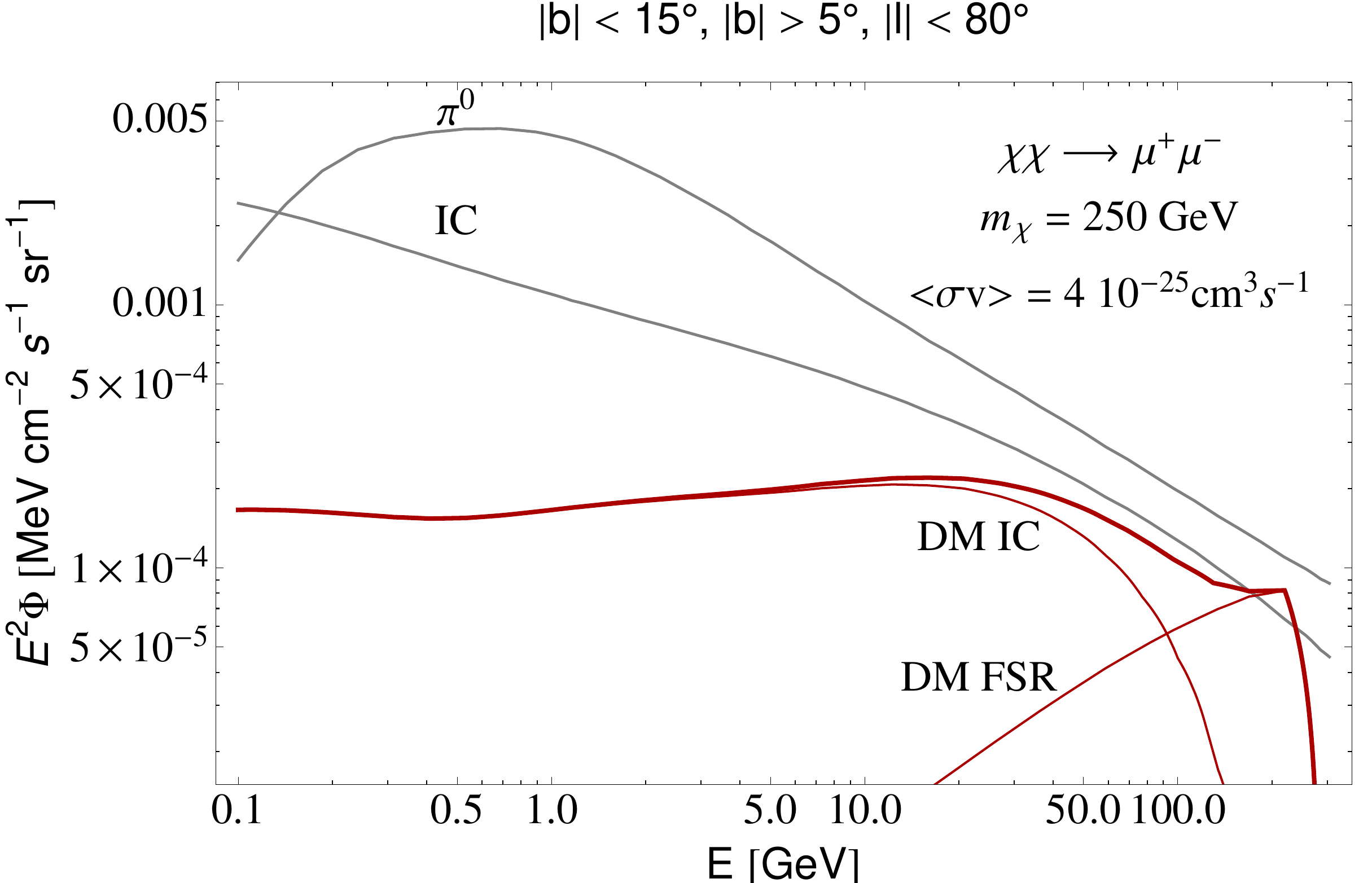} \\
\includegraphics[width=0.45\textwidth]{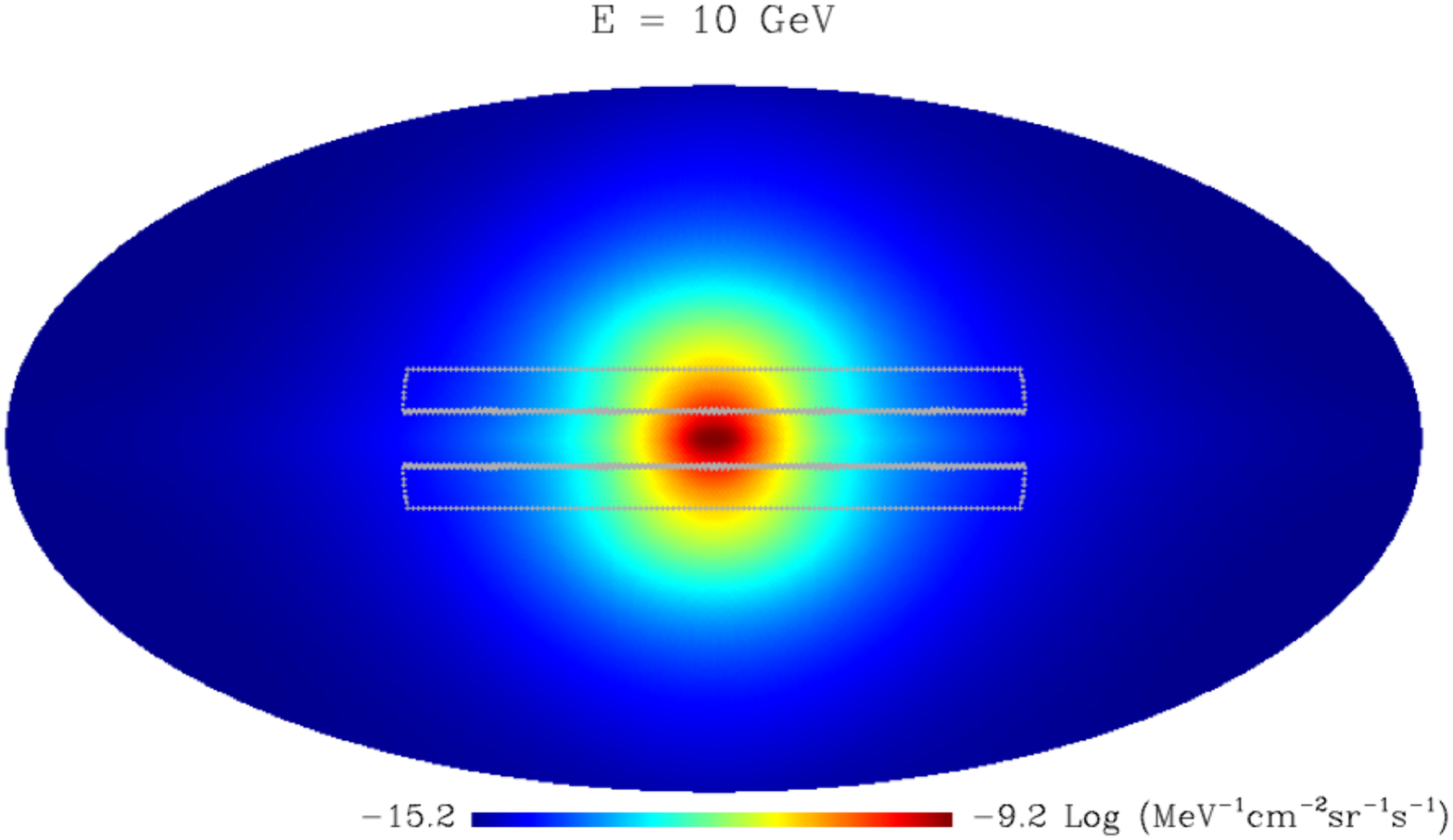}&
\includegraphics[width=0.42\textwidth]{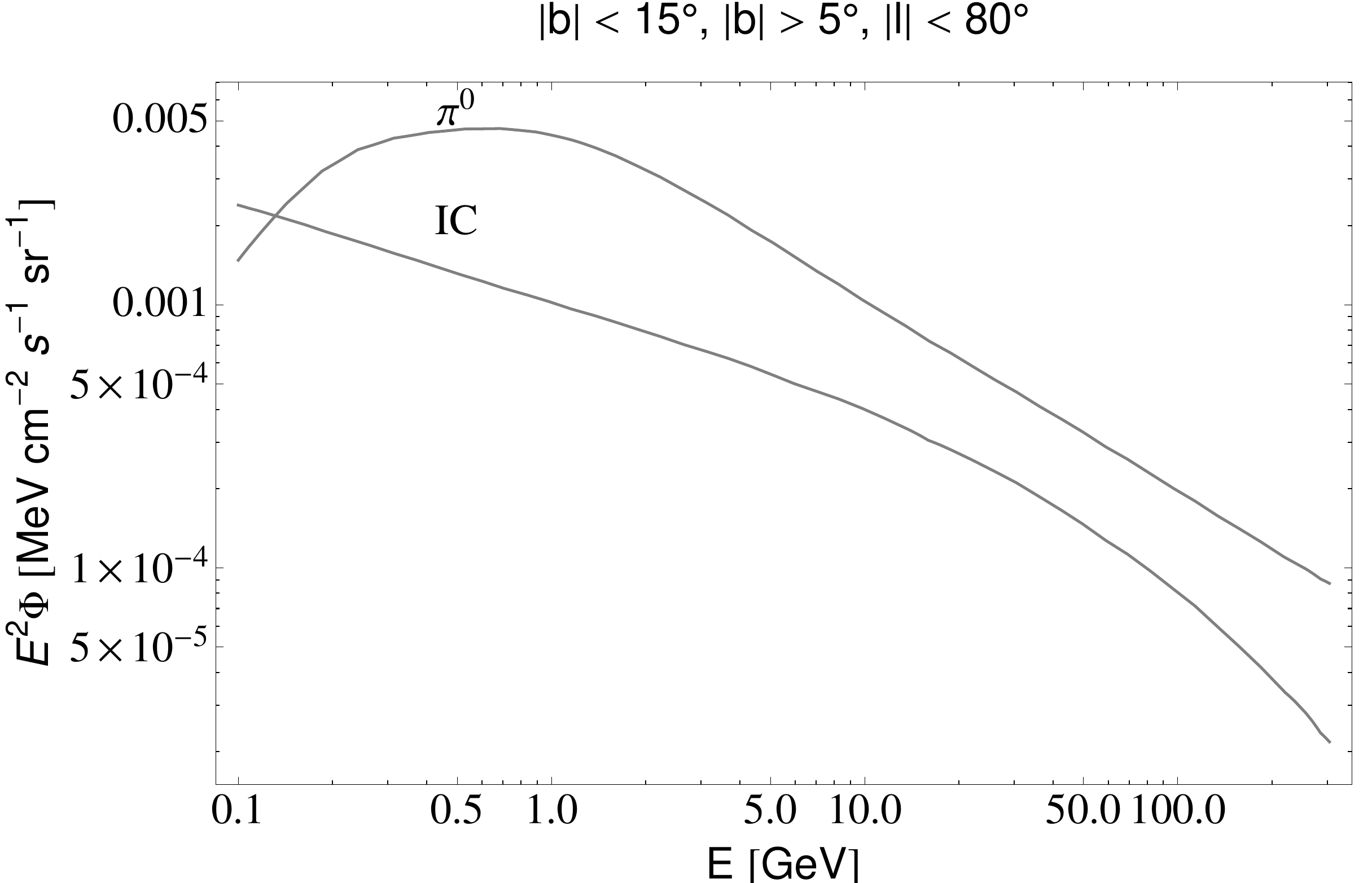}
\end{array}$
\end{center}
\vspace{-0.5cm}
\caption{\emph{Upper panel}: Spatial (left) and spectral (right) distribution of gamma rays originating from the annihilation of a 250 GeV WIMP into $b{\bar b}$. The left figure shows the expected intensity at E=10 GeV for the full sky in Galactic coordinates. A NFW profile is assumed for the DM halo and a value of $\left< \sigma_A v \right> =4\times10^{-25}$cm$^3$s$^{-1}$ for the DM annihilation cross section. For comparison purposes typical spectra of the astrophysical emission from $\pi ^0$ decay and Inverse Compton (IC) scattering  are displayed in the right figure. The map also shows the boundaries of the region used to plot the average spectra of the right panel, and which we will use for the analysis described in this work.  \emph{Central panel}: Same for a 250 GeV WIMP annihilating into $\mu ^+ \mu^-$. The contribution from IC and from Final State Radiation (FSR) are shown separately in the spectrum and are superimposed in the spatial distribution. 
\emph{Lower panel}: Spatial (left) and spectral (right) distribution of  the IC emission of an astrophysical CR source population distributed uniformly in Galactocentric radius within 1 kpc from the Galactic Center and with a  scale height of 200 pc.   \label{dmmaps} }
\end{figure*}

\section{Modeling of the high-energy Galactic diffuse emission}\label{diffusemodeling}
The Galactic diffuse gamma-ray emission is produced by the interaction of energetic cosmic-ray (CR) electrons and nucleons with the interstellar gas and radiation field. Its main components are photons from the {\em decay of neutral pions} produced in the interaction of the CR nucleons with the interstellar gas, {\em bremsstrahlung} of the CR electron (CRE) population on the interstellar gas and their {\em inverse Compton} (IC) scattering off the Interstellar Radiation Field (ISRF).  Efficient modeling of the diffuse gamma-ray emission needs an accurate description of both the interstellar gas and radiation targets as well as the distribution of CRs in the Galaxy.

The gas in the Interstellar Medium (ISM) consists mostly of {\em atomic} hydrogen (H~{\sc i}) and, to a lesser extent,  {\em molecular} hydrogen (H$_2$) which is concentrated along the Galactic Plane.   {\em Ionized} (H~{\sc ii}) hydrogen is subdominant, although it has a larger vertical scale (see \cite{2012arXiv1202.4039T,2004ASSL..304..279M} for more details). Helium (He) is also important, being  $\sim 25$\% by mass of the ISM, and its distribution is assumed to  follow that of interstellar hydrogen \citep{2012arXiv1202.4039T}.
Velocity resolved radio surveys of the 21 cm hyperfine structure transition of H~{\sc i} and corresponding surveys of the 2.6 mm CO J($1\rightarrow 0$) transition (using the CO density as a proxy for the H$_2$ density) are used to build maps of the interstellar gas in different annuli  \citep{2012arXiv1202.4039T} providing effectively a 3D\footnote{More precisely the model is only pseudo 3D due to the near-far ambiguity in the inner Galaxy  \citep{2012arXiv1202.4039T}.} model of the gas distribution in the Galaxy. The conversion factors $X_{CO}$ between CO line intensity and H$_2$ column density have been observed to vary throughout the Galaxy \citep{2010ApJ...710..133A}. 
Total gas column density estimated from E(B-V) visual reddening maps \citep{1998ApJ...500..525S} has been shown to be complementary to the one estimated from H~{\sc i} and CO surveys combined \citep{2005Sci...307.1292G}.  As described in \cite{2012arXiv1202.4039T}, we take this into account by correcting the gas column density for each line of sight according to the value derived from the E(B-V) map, except in the regions of high extinction (see Sec.~\ref{gasmaps}).

We use a 2D+1 cylindrically symmetric model (2 spatial dimensions and the frequency dimension) of the ISRF, computed based on a model of the radiation emission of stellar populations and further reprocessing in the Galactic dust \citep{2006ApJ...640L.155M}.

We use the  \texttt{GALPROP} code \citep{2000ApJ...537..763S} v54, to calculate the propagation and distribution of CRs in the Galaxy. 

The code is further used to create sky maps of the expected gamma-ray emission from the interactions of the CRs with the ISM and ISRF based on the  models of the gas and radiation targets described above. \texttt{GALPROP} approximates the CR propagation by a diffusion process into a cylindrical diffusion zone of half-height $z_h$ and radius $R_h$.  
CREs and {nuclei} 
are injected  by a parametrized distribution of CR sources. Energy losses, production of secondary particles in interactions and reacceleration of CRs in the ISM are taken into account (for details see  \citealp{2000ApJ...537..763S}).  Several important parameters enter the \texttt{GALPROP} modeling: the distribution of CR sources, the half-height of the diffusive halo $z_h$, the radial extent of the halo $R_h$,  the nucleon and electron injection spectrum, the normalization of the diffusion coefficient $D_0$, the rigidity dependence of the diffusion coefficient $\delta$ ($D(\rho)=\beta D_0 (\rho/\rho_0)^{-\delta}$ with $\rho_0=4$ GV being the reference rigidity and $\beta=v/c$) the Alfv{\'e}n speed $v_A$ (parametrizing the strength of re-acceleration of CRs in the ISM via Alfv{\'e}n waves) and the velocity of the Galactic winds perpendicular to the Galactic Plane $V_c$. Following \cite{2012arXiv1202.4039T} we will parameterize the nucleon injection spectrum as a broken power law in rigidity with $\gamma_{p,1}, \gamma_{p,2}$ the index of the spectrum before and after the break respectively and $\rho_{br, p}$ the break rigidity. Similarly, the electron injection spectrum is parametrized as a double broken power law with $\gamma_{e,1}, \gamma_{e,2},  \gamma_{e,3},$ the index of the spectrum in the three rigidity zones and  $\rho_{br, e,1}, \rho_{br, e,2}$ the two breaks.

\begin{table*}[t]
\begin{center}
\begin{tabular}{|c|c|c|c|c|c|c| }
\hline
\textbf{Parameter}  & \multicolumn{6}{|c|}{  \textbf{Value} } \\
\hline \hline
 Halo Height $z_h$(kpc)&  2 & 4  & 6   & 8  & 10  & 15 \\
 Diffusion Coefficient $D_0$ (cm$^2$s$^{-1}$)&  $2.7\times10^{28}$  & $5.3\times10^{28}$  &  $7.1\times10^{28}$  &  $8.3\times10^{28}$  &  $9.4\times10^{28}$  &  $1.0\times10^{29}$\\
 Diffusion Index $\delta$ & 0.33 & 0.33 & 0.33  & 0.33  & 0.33  & 0.33  \\
 Alfven Velocity $v_A$  (km s$^{-1}$) & 35.0 & 33.5 & 31.1   & 29.5  & 28.6  & 26.3 \\
 Nucleon Injection Index (Low)  $\gamma_{p,1}$ & 1.86 &  1.88 & 1.90  & 1.92  & 1.94  & 1.96\\
 Nucleon Injection Index (High) $\gamma_{p,2}$ & 2.39 &  2.39 & 2.39  & 2.39  & 2.39  & 2.39\\
 Nucleon  break rigidity $\rho_{br, p}$(GV) & 11.5 & 11.5 & 11.5  & 11.5  & 11.5  & 11.5\\
\hline
\end{tabular}
\caption{CR diffusion   parameters from \cite{2012arXiv1202.4039T} used in this work.}
\label{pdiffusion}
\end{center}
\end{table*}

We use in the following the results and formalism of  \cite{2012arXiv1202.4039T} and we briefly summarize here the approach and results presented there. The reader is referred to  \cite{2012arXiv1202.4039T} for more details and thorough discussion.  
In that work a grid of models is considered with 4 values of $z_h$, { 2 values of $R_h$} and 4 different models of CR Source Distributions (CRSDs). The CRSDs  are set to correspond to the incompletely determined distributions of Supernova Remnants (SNR), or tracers of star formation and collapse (pulsars, OB stars) \cite{1998ApJ...504..761C,2004A&A...422..545Y}.  {For each of these 32 models 4 different assumptions on the column density of the hydrogen gas derived from its tracers are made for a total of 128 models.}  For each of the models, a fit of the model prediction to the local intensity of different CR nuclei and the B/C ratio is performed in order to fix the parameters ($D_0,v_A, \gamma_{p,1}, \gamma_{p,2}, \rho_{br}$).  
Thus, the CR fit  provides the injection spectrum for nuclei, the Alfv{\'e}n speed and the relation $(z_h, D_0)$ for different values of $z_h$.  In a second step they determine also the electron injection spectrum from a fit of the model to the local spectrum of CRE using the diffusion parameters obtained in the first fit. However, we are not going to use the result of the second step, since we will instead fit the electron spectrum from  gamma-ray data. As the last step in \cite{2012arXiv1202.4039T}  an all-sky fit to the {\it Fermi}-LAT gamma-ray data is then performed to find the remaining parameters like the $X_{CO}$ factors\footnote{The procedure is iterated few times in order to consider the feedback of the renormalized $X_{CO}$ factors in the propagation of CREs.}.
The ISRF normalizations in various regions of the sky are also left as parameters free to vary in the fit, to account for possible uncertainties in the ISRF itself and  the CRE distribution. In our analysis, we will also consider the ISRF uncertainties in more detail in Sec. \ref{discussion}.

Thus, we use only the results from the first step of the analysis in \cite{2012arXiv1202.4039T} described above, but then allow for more freedom in certain parameters governing the CR distribution and astrophysical diffuse emission and constrain these parameters by fitting the models to the LAT gamma-ray data.
 Compared to \cite{2012arXiv1202.4039T}, the main difference in our analysis is the use of a free CRSD whose parameters will be determined from the fit to the LAT data, as opposed to the 4 fixed CRSDs explored in \cite{2012arXiv1202.4039T}.  The procedure  which we employ to obtain the best fit CRSD is described in detail in Sec. \ref{main}.  Another important difference is that we will keep the CRE source distribution and proton source distribution (which we will refer to as $e$CRSD and $p$CRSD) separate. This is justified as we don't know a priori if the bulk of CR protons and electrons is injected by the same class of sources. 
We report in Table \ref{pdiffusion} the CR diffusion and injection parameters for different values of $z_h$ taken from \cite{2012arXiv1202.4039T} which we will use in the following\footnote{More precisely, the $z_h$=2, 15 cases are not reported in \cite{2012arXiv1202.4039T}. The values used here for $z_h=2,15$ are, however, obtained in the same way as the other $z_h$ cases.}.
Note that the diffusion parameters in principle depend on the CRSD, but the dependence is weak and will be neglected in the following. There is also slight dependence on the parameters used to produce the gas maps (see Sec. \ref{maps}) and the assumed $X_{CO}$ distribution, which is also weak and will be neglected as well.
Besides the free CRSDs and the scan over different values of $z_h$ we will also scan electron injection spectra by varying the index  $\gamma_{e,2}$ while we will fix  $\gamma_{e,1}=1.6$, $\gamma_{e,3}=4$ \citep{2012arXiv1202.4039T},   $\rho_{br, e,1}=2500$ MV and  $\rho_{br, e,3}=2.2$ TV.  The last two parameters are left free to vary in the analysis performed in \cite{2012arXiv1202.4039T} although the fitted values typically differ by less than 20\% for $\rho_{br, e,1}$ and less than 10\% $\rho_{br, e,3}$  with respect to the values we report above.

As a sanity check it should be also verified a posteriori that the flux and spectrum of local protons obtained after the gamma-ray fit are consistent
with the experimentally observed ones, since the $p$CRSD fit could, potentially, affect them, and this would make inconsistent the use of the initial diffusion parameters, based on the observed proton spectrum. We, indeed, found that the proton spectrum after the second step is fully consistent with the observed one.
We also verified that the normalization of the electron flux given by the gamma-ray fit is consistent with the observed one. This, though, 
is not strictly required for the consistency of the approach, since the CRE spectrum is not used to determine the diffusion parameters.

We also note that the nucleon injection spectra and the diffusion parameters are solely determined by fitting the local CR density, and neglecting the effect of DM. Injection of large quantities of nucleons  from DM near the Galactic Center, that could alter the local abundances of CR nuclei (like the proton spectrum and the local B/C ratio, used to perform the fit)  are in fact strongly excluded by anti-matter measurements. The anti-proton fraction, for example,  recently measured by the PAMELA experiment up to $\sim$100 GeV  \citep{2009PhRvL.102e1101A} is about $10^{-4}$ above 10 GeV and thus constrains the DM contribution to be below this value.

A drawback of using CR data is that they are usually affected by large systematics (for example, in the energy scale or in the solar modulation correction), so that the errors on the inferred diffusion parameters are larger than the statistical errors obtained from the fit (see \cite{2011ApJ...729..106T} for a recent attempt to take into account these systematic effects).  
We checked that even quite large variations in these parameters affect our results only weakly. A more detailed discussion is deferred to Sec. \ref{discussion}.

\subsection{Limitations of the model}\label{sec:caveats}
The model described above represents well the gamma-ray sky, although various residuals ({at a $\sim 30\%$ level \citep{2012arXiv1202.4039T}}), both at small and large scales, remain. 
These residuals can be ascribed to various limitations of the models: i) imperfections in the modeling of gas and ISRF components, ii) simplified assumptions 
in the propagation set-up (e.g. assumption of isotropy and homogeneity of the diffusion coefficient),  iii) unresolved point sources, which are expected to contribute to the diffuse emission at a level of $10\%$ \citep{2007Ap&SS.309...35S}, and have not been taken into account in the modeling in \cite{2012arXiv1202.4039T},  iv) missing structures like Loop I \citep{2009arXiv0912.3478C} or the Galactic Bubbles/Lobes \citep{2010ApJ...724.1044S}. 

We will deal with these limitations in various ways. Structures like Loop I and the Galactic Bubbles appear mainly at high Galactic latitudes and their effects on the fitting can be limited using a Region of Interest (ROI) with limited extent in Galactic latitude.  In the following we will consider a ROI in Galactic latitude, $b$, of $5^{\circ} \leq |b|\leq 15^{\circ}$, and Galactic longitude, $l$, $|l|\leq 80^{\circ}$, see Sec. \ref{sec:ROI}. 
As for small scale residuals, we believe they are due to imperfections in the gas maps. In order to quantify their effect on the constraints of the DM properties, we will calculate the likelihood for several different assumptions on the gas total column density.

Despite the various choices described above and  the large freedom we leave in the model (CRSDs, $z_h$, index of the electron injection spectrum),
we still see residuals in our ROI at the $\pm 30\%$ level and at $\agt 3~\sigma$ significance (see the figures in Sec.\ref{results}).
Positive residuals, in particular, appear in various places in the ROI, especially in connection with the low Galactic latitude extension of the Lobes and Loop I.
Residuals at the same level appear also when we perform a fit in a high latitude ($|b|\agt40^\circ$) control region, where DM is not expected to contribute significantly, and thus seem to indicate the general level of accuracy achievable with the present modeling of the diffuse emission.
Note that the residuals related  to the Lobes and Loop I do not appear in the official {\it Fermi}-LAT diffuse model\footnote{For a description see http://fermi.gsfc.nasa.gov/ssc/data/access/lat/BackgroundModels.html.} since there they are explicitly modeled through the use of patches.
Since the residuals do not seem obviously related to DM, as it would be, for example, in the case of a single strong positive residual near the Galactic Center, we thus decide to focus in the following on setting limits on the possible DM signal, rather than \emph{searching} for a DM signal.
Given the presence of the residuals like those described above, we also decide to quote more generous limits at the 3 and 5 sigma level, as well as conservative limits without assumptions over the astrophysical background.
A dedicated \emph{search} for a DM signal will be reported in a forthcoming paper, where data in the Galactic Plane and at low energies ($\alt 1$ GeV) also will be exploited  in order to help separating a real DM signal from other astrophysical processes which may be responsible for the above residuals.

\section{Outline of the Approach to set  DM limits} \label{outline}

Having a parameterized model of the astrophysical Galactic diffuse emission as described above, we could imagine exploring potential DM contributions by adding the DM component in the fit and performing a global joint fit of DM and the astrophysical model parameters.
In practice, however, it is at the moment computationally challenging 
to perform such a large global fit. Thus, some additional simplifying assumption must be made.
As already outlined in the previous section, the main simplification which we will adopt (which is also used in \cite{2012arXiv1202.4039T}) is the splitting of the fit in two separate steps constraining some parameters based on measurements of the local CR intensities while constraining other parameters based on {\it Fermi} LAT data. 
For the first step  
we rely entirely on \cite{2012arXiv1202.4039T} and we use the 6 different diffusion models (for 6 different $z_h$) whose parameters are summarized in Table \ref{pdiffusion}. We then use only gamma-ray data to perform the fit 
for the CRSDs, $z_h$, the electron index, gas maps with different column densities of the interstellar gas, and DM.
Some additional parameters which enter the gamma-ray fit are further introduced in Sec.~\ref{sec:linearpar}.

Our aim is to constrain the DM properties  
and treat the parameters of the astrophysical diffuse gamma-ray background as nuisance parameters. Those parameters are typically correlated 
with the assumed DM content and it is thus important to consider them since they affect directly the DM fit. 
It is clear, for example, that the CRSD should have a large influence on the fit of the DM component.
This can be seen from Figure  \ref{dmmaps} which shows that the gamma-ray signal produced by DM is somewhat degenerate with the IC signal from  CR sources placed in the inner Galaxy. 
Besides small morphological differences they mainly differ in the energy spectrum, which, however, is quite model dependent in the DM case.  
To explore the effect of uncertainties in the foreground modeling we use the LAT data to fit the CRSD for both the nuclei and CREs, as well as the CRE injection spectrum which directly affects the IC component.
It is important to stress that to constrain DM in a self-consistent way, we will fit the CR distributions and DM at the same time, in order to take into account the degeneracy between the two. 
Fortunately, in our ROI, above and below the inner Galaxy, and around few GeV in energy, the $\pi^0$ component, which has different morphology and is not degenerate with the DM signal, is dominant over IC and bremsstrahlung by about a factor $4\--5$. So, we expect approximately the same factor in improvement in DM constraints with respect to the fits that obtain limits without modeling this background. We will see in Sec. \ref{nobkg} and \ref{results} that this expectation approximately holds.

With the general approach above we set DM limits using the profile likelihood method (outlined in Sec.~\ref{profilelikelihood}). 
Besides  the approach above,  we will also quote conservative upper limits using  the data only (i.e.  without performing any modeling of the astrophysical background). These conservative limits are along the lines of the  work of \cite{2010JCAP...03..014P,2010NuPhB.840..284C}, which use a similar approach to set DM limits based on the first year of {\it Fermi} LAT data.

\section{Maps}\label{maps} 

\subsection{DM maps}
The template maps used in the fits to model the DM contribution depend on the assumed DM distribution and the assumed annihilation/decay channel. Numerical simulations of the Milky Way-scale halos indicate a smooth distribution that contains a large number of subhalos \citep{2007ApJ...667..859D,2008MNRAS.391.1685S}. 
The gamma-ray signal from the subhalo population is expected to dominate in the region of the outer halo, while  in the inner $\lsi 20^\circ$ region of the Galaxy, its contribution is expected to be subdominant  \citep{2007ApJ...657..262D,2008Natur.456...73S,2011PhRvD..83b3518P}. In our ROI the subhalo contribution therefore should be mild and we conservatively consider only the smooth component in this work. We parametrize the  smooth DM density $\rho$ with a NFW spatial profile \citep{1996ApJ...462..563N} and a cored (isothermal-sphere) profile \citep{1991MNRAS.249..523B,1980ApJS...44...73B}: 
\begin{eqnarray}
  {\rm NFW}\! :  \,\,\, \rho(r) &=&\rho_0  \left(   1+\frac{R_\odot}{R_s}  \right)^2  \frac{1}{\frac{r}{R_\odot}  \left(   1+\frac{r}{R_s}  \right)^2}    \\
 {\rm Isothermal}\!  :   \,\,\, \rho(r) &=& \rho_0 \ \frac{R_\odot^2+R_c^2}{r^2+R_c^2}. 
\end{eqnarray}
These are traditional benchmark choices, as NFW is motivated by N-body simulations, while cored profiles are instead motivated by the observations of  rotation curves of galaxies and are also found in simulations of a Milky Way-scale halos involving baryons \citep{2012ApJ...744L...9M}. The Einasto profile \citep{2006AJ....132.2685M,2010MNRAS.402...21N} is emerging as a better fit to more recent numerical simulations, but for brevity we do not consider it here. It is expected that this profile should lead to DM limits stronger by $\sim 30\%$ in our ROI, with respect to a choice of a NFW profile \citep{2010NuPhB.840..284C}.
The main uncertainty in the DM halo profile comes from the poorly known (and modeled) baryonic effects.
Indeed, with our choice of NFW and Isothermal profiles, our aim is to roughly bracket the uncertainties expected
from the DM profile.
For the local density of DM we take the value of \mbox{$\rho_0=0.43$ GeV cm$^{-3}$}  \citep{2010A&A...523A..83S} \footnote{The measurement has a 
typical associated error bar of $\pm$0.1 GeV cm$^{-3}$ and a possible spread up to 0.2-0.7 GeV cm$^{-3}$ \citep{2010A&A...523A..83S,2011JCAP...03..051C}.}, and the scale radius is assumed to be $R_s=$ 20 kpc (NFW) and $R_c=$ 2.8 kpc (isothermal profile). The actual choice of the DM density profile does not have a major effect on our limits (see Sec. \ref{results}) as we do not consider the central few degrees of the Galaxy (where these distributions differ the most). A choice of a more extended core of $\sim 5$ kpc seems possible, although less favored by  data \citep{1998APh.....9..137B} 
 (see also   \cite{2010JCAP...08..004C,2010A&A...509A..25W,2011JCAP...11..029I} for further discussions on the $\rho_0$ and DM profile uncertainties).
 With this choice our limits would worsen by a factor of $\lsi ~2$. We also set the distance of the solar system from the center of the Galaxy  to the value $R_\odot=$ 8.5 kpc \citep{1986MNRAS.221.1023K}.

For the annihilation/decay spectra we consider three channels with distinctly different signatures: annihilation/decay into the $b{\bar b}$ channel,  into $\mu ^+ \mu^-$, and into $\tau^+\tau^-$ . In the first case gamma rays are produced through hadronization and pion decay. The resulting spectra are similar for all channels in which DM annihilations/decays produce heavy quarks and gauge bosons in the energy range considered here and is therefore representative for a large set of WIMP particle physics models.
The choice of annihilation/decay into leptonic channels, provided by the second and third scenarios, is motivated by the PAMELA  positron fraction \citep{2009Natur.458..607A} and the {\it Fermi} LAT  electrons plus positrons \citep{2009PhRvL.102r1101A}  measurements (for interpretation of these measurements in terms of a DM signal see e.g. \citealp{2009APh....32..140G,2010NuPhB.831..178M,2009PhRvL.103c1103B}). In this case, gamma rays are dominantly produced through radiative processes of electrons, as well as through the Final State Radiation (FSR).  

We produce the DM maps with a version of \texttt{GALPROP}  slightly modified to implement custom DM profiles and injection spectra. For the prompt photons case \texttt{GALPROP} integrates along the line of sight the DM gamma-ray emissivity given by $Q_{\gamma}(r,E)=\rho^2 \left< \sigma_A v \right> /2m_{\chi}^2 \times dN_{\gamma}/dE$ in the annihilation case and by $Q_{\gamma}(r,E)=\rho \ \! \Gamma_D /m_{\chi} \times dN_{\gamma}/dE$ in the decay case,  $m_{\chi}$ being the DM particle mass, $dN_{\gamma}/dE$ the gamma-ray annihilation/decay spectrum, $\left< \sigma_A v \right> $ the thermally averaged DM annihilation cross section  and $\Gamma_D=1/\tau$, the DM decay rate (i.e. the inverse of the DM lifetime).
In the cases where the propagation of electrons produced in DM annihilation is relevant to the resulting gamma-ray emission, template maps are produced including the IC emission from DM annihilation/decay-generated  CREs, which have been propagated using the relevant set of propagation parameters. 
An example of the resulting DM maps at 10 GeV  
for two DM models (annihilation to $b\bar{b}$ and $\mu ^+\mu^-$ channels) is shown in Figure \ref{dmmaps}, for a DM mass of 250 GeV.

We calculate the DM injection spectrum of electrons and gamma rays by using  {the \texttt{PPPC4DMID} tool} described in \cite{2011JCAP...03..051C}. This package provides interpolating functions calculated from a simulation of DM annihilation/decay with the \texttt{PYTHIA} \citep{2008CoPhC.178..852S} event generator. It has recently been shown that for DM candidates with masses above the electroweak scale, the standard predictions for the annihilation/decay spectra are altered when account is taken of the production of electroweak gauge bosons from the FSR \citep{2009PhRvD..80l3533K,2011JCAP...03..019C,2011JCAP...06..018C}. We include these electro-weak corrections here, following \cite{2011JCAP...03..051C}, even though the effect on DM limits is marginal for our choice of energy range.

\subsection{Gamma-ray emission from CR interactions with interstellar gas}\label{gasmaps}

The astrophysical diffuse emission predicted by \texttt{GALPROP} depends on assumptions about the distribution and column density of the interstellar gas entering the model.  
A significant uncertainty is related to the total gas column density due to the presence of dark gas. The dark gas contribution is estimated 
using the Schlegel, Finkbeiner, Davis (SFD) \cite{1998ApJ...500..525S} E(B-V) dust reddening map as a tracer of the total integrated gas column density (dark + radio-visible). The intensity of the velocity resolved H~{\sc i} maps derived from the 21-cm survey data is then rescaled by the ratio between the total and radio-visible column densities to account for dark gas. This rescaling factor depends on 
the ratio between the dust and gas column densities and thus on the dust to H~{\sc i} ratio (d2HI) and the dust to CO ratio (d2CO).  In \cite{2012arXiv1202.4039T} the dust to H~{\sc i} and CO gas ratios are fixed through a regression procedure of the SFD map to the radio-derived gas maps (see \cite{2012arXiv1202.4039T} for more details) which yields d2HI=0.0137 $\times 10^{-20}$ mag cm$^2$ and d2CO=0.0458 mag (K km s$^{-1}$)$^{-1}$. Here we will instead explore different values of the dust to gas ratios, to consider the possibility that this quantity is different in our ROI with respect to the all-sky derived value of  \cite{2012arXiv1202.4039T} and thus, possibly, improve the residuals, especially at small scale, as discussed in Sec. \ref{sec:caveats}. In particular we will use 6 values of d2HI equally spaced in the range $(0.0120-0.0170)$ $\times 10^{-20}$ mag cm$^2$.  
The d2CO ratio will be instead fixed to the value 0.04 mag (K km s$^{-1}$)$^{-1}$. This is justified in the light of the fact that above $\pm5^\circ$ of Galactic latitude, where we perform the fit, there is very little CO and we are thus not very sensitive to this parameter.
Nevertheless, we checked the results for different values  of  d2CO in the range $0.03-0.06$ mag (K km s$^{-1}$)$^{-1}$ and found no appreciable change in the results.

The E(B-V) map is not a reliable tracer of total column density when the dust reddening becomes very high. Thus a cut in E(B-V) needs to be employed to exclude the regions of high Galactic extinction. In \cite{2012arXiv1202.4039T} a value of E(B-V)$<5$ mag is found to be adequate and we will use it in the following. This cut excludes   from the dark gas correction a narrow strip of few degrees along the Galactic Plane, which is a region outside our ROI.

The spin temperature of the H~{\sc i}, $T_S$ (see \cite{2012arXiv1202.4039T} for the detailed definition), used to extract the H~{\sc i} maps from the radio data is also not very well known and introduces further uncertainties.  However, since in \cite{2012arXiv1202.4039T} the total column  density of the ISM is estimated from dust, the effect of $T_S$ is typically subdominant with respect to the dark gas correction. We will thus not consider it and fix $T_S$ to the typical value $T_S=150$ K \citep{2012arXiv1202.4039T}.

\section{Data selection and region of interest}\label{sec:ROI}
 
We use 24 months LAT data starting from 2008 August 5 to 2010 July 31,  in the energy range between 1 GeV and 100 GeV. However, we use energies up to 400 GeV when deriving DM limits with no assumption on the astrophysical background, see next section.  
As the data above 100 GeV have poor statistics they have little weight when performing the fit to the gamma-ray data to determine the background model. Instead, for the no-background limits, where no fitting is performed, high energy points are relevant to determine the DM limits only for the hard FSR spectrum. 
We use only events classified as gamma rays in the P7CLEAN event selection and the corresponding \verb"P7CLEAN_V6" instrument response functions\footnote{http://fermi.gsfc.nasa.gov/ssc/}. The data have been extracted and processed with the {\it Fermi} tools as described in \cite{2012arXiv1202.4039T}. 
In order to minimize the contribution from the very bright Earth limb, we apply a maximum zenith angle cut of $90^\circ$. In addition, we also limit our data set to include only photons with an incidence angle from the instrument $z$-axis of $< 72^\circ$.
The events are divided into 5 logarithmically spaced energy bins. The total number of photons in our ROI is $\sim 350000$.
We use a HEALPix\footnote{http://healpix.jpl.nasa.gov}~\citep{2005ApJ...622..759G} \verb"nside"=64 pixelization scheme for the spatial binning, corresponding to a bin size of approximately $0.9$ deg $\times ~0.9$ deg.  We further  mask the point sources from the 1FGL source catalog \citep{2010ApJS..188..405A}  to limit the impact of bright point sources on the fit.  We mask all pixels within 1 deg of the location of the point sources.  With this choice we remove $\sim 25\%$ of the photons, which leaves us with $\sim 270000$. The 2FGL source catalog \citep{2012ApJS..199...31N} contains some further weak point sources which we do not mask to not impact dramatically the size of our ROI.  The effect of masking 2FGL sources is further considered in Sec. \ref{results}.     
Since we  focus on the Galactic halo we choose a ROI limited by $\pm{15^\circ}$ in Galactic latitude and $\pm{80^\circ}$ in longitude, to focus on the region where the S/N ratio for DM is the highest \citep{2008APh....29..380S}, and to minimize the effect of the high latitude structures like Loop I or the Galactic Lobes. This ROI also excludes the Outer Galaxy  where the DM signal should be lower. 
Furthermore, we mask the region $|b|~\lsi 5$ deg along the Galactic Plane, in order to reduce the uncertainty due to the modeling of the astrophysical and DM emission profiles discussed above.

\section{DM limits with no assumption on the astrophysical background}\label{nobkg}
To set  DM limits with no assumption on the astrophysical background we first convolve a given DM model with the exposure maps and Point-Spread Function (PSF) to obtain the counts expected from DM annihilation. The \texttt{GaRDiAn} (Gamma-Ray DIffuse ANalysis) software package \citep{2012arXiv1202.4039T,Ackermann:2009zz} is used for this processing. The expected counts are then compared with the observed counts in our ROI and the upper limit is set to the \emph{minimum} DM normalization which gives counts in excess of the observed ones in at least one bin.
More precisely, we set $3\sigma$ upper limits given by the requirement  $n_{i DM}-3\sqrt{n_{i DM}} > n_i$ \citep{2011EPJC...71.1554C}, where $n_{i DM}$ is the expected number of counts from  DM in the bin $i$ and $n_i$ the actual observed number of counts. It should be noted that the formula assumes a Gaussian model for the fluctuations, which is a good approximation given the large bin size and the number of counts per bin we use in this case (see below). 
The large Poisson noise present especially at high ($>10$ GeV) energies due to the limited number of counts per pixel,  affects the  limits for DM masses above 100 GeV, weakening them somewhat.  To reduce the Poisson noise, \emph{only} for the present case of no background modeling we choose a larger pixel size so to increase the number of counts per pixel. However, a very large pixel size would  wash out the DM signal, diluting it in large regions, again weakening the limits. 
We chose the case with a pixel size of about $7$$^\circ$ $\times ~7$$^\circ$ (\verb"nside"=8) since it gives a reasonable compromise between the two competing factors\footnote{The mask is always defined (and applied) at $\texttt{nside}=64$. After applying the mask the data (and the models) are downgraded to the larger pixel size.}. In this way limits typically improve by a factor of a few with respect to the case \verb"nside"=64. Limits for DM masses below 100 GeV, instead, are only very weakly affected by the choice of the pixel size in the range $1^\circ\--7$$^\circ$. 
Finally, again only for the present case of no modeling of the background, we use an extended energy range up to 400 GeV. This, in practice, is important only for the $\mu^+\mu^-$ case for masses above 100 GeV and  when we consider FSR only (since the $\mu^+\mu^-$ FSR annihilation spectrum is peaked near the energy corresponding to the DM mass and thus  can be constrained only by using higher-energy data). For the other cases, instead, there is always significant gamma-ray emission below 100 GeV, either from prompt or IC photons and the extended energy range does not affect the limits appreciably.
  
The limits derived from this analysis are discussed in Sec.\ref{results}.  These constraints are about a factor of 5 worse than those obtained with a  modeling of the background (see next section), which is in agreement with the estimate made in Sec. \ref{outline}.

\begin{figure*}[tp] 
\begin{center}$
\begin{array}{cc}
\includegraphics[width=0.45\textwidth]{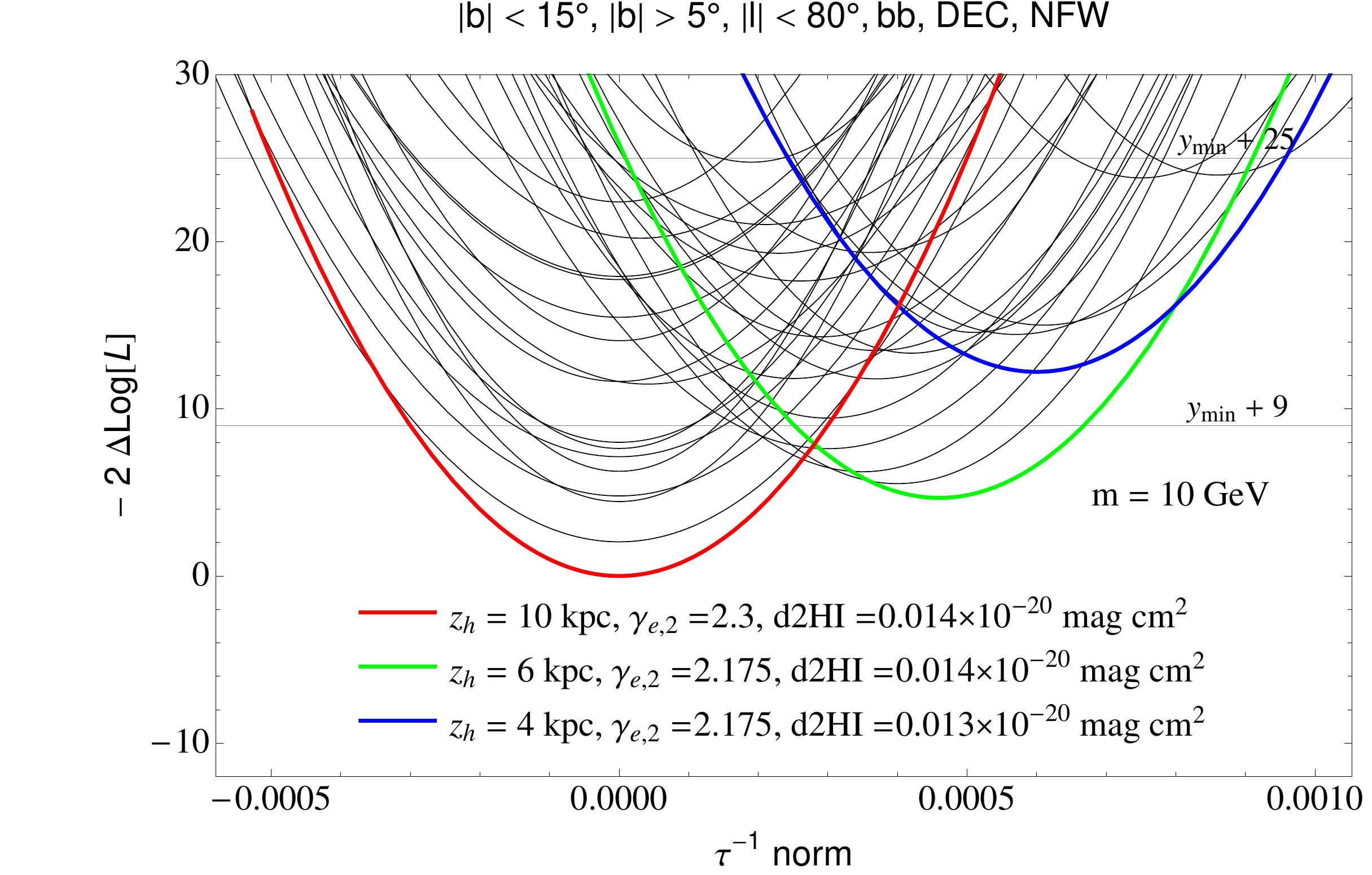}&
\includegraphics[width=0.45\textwidth]{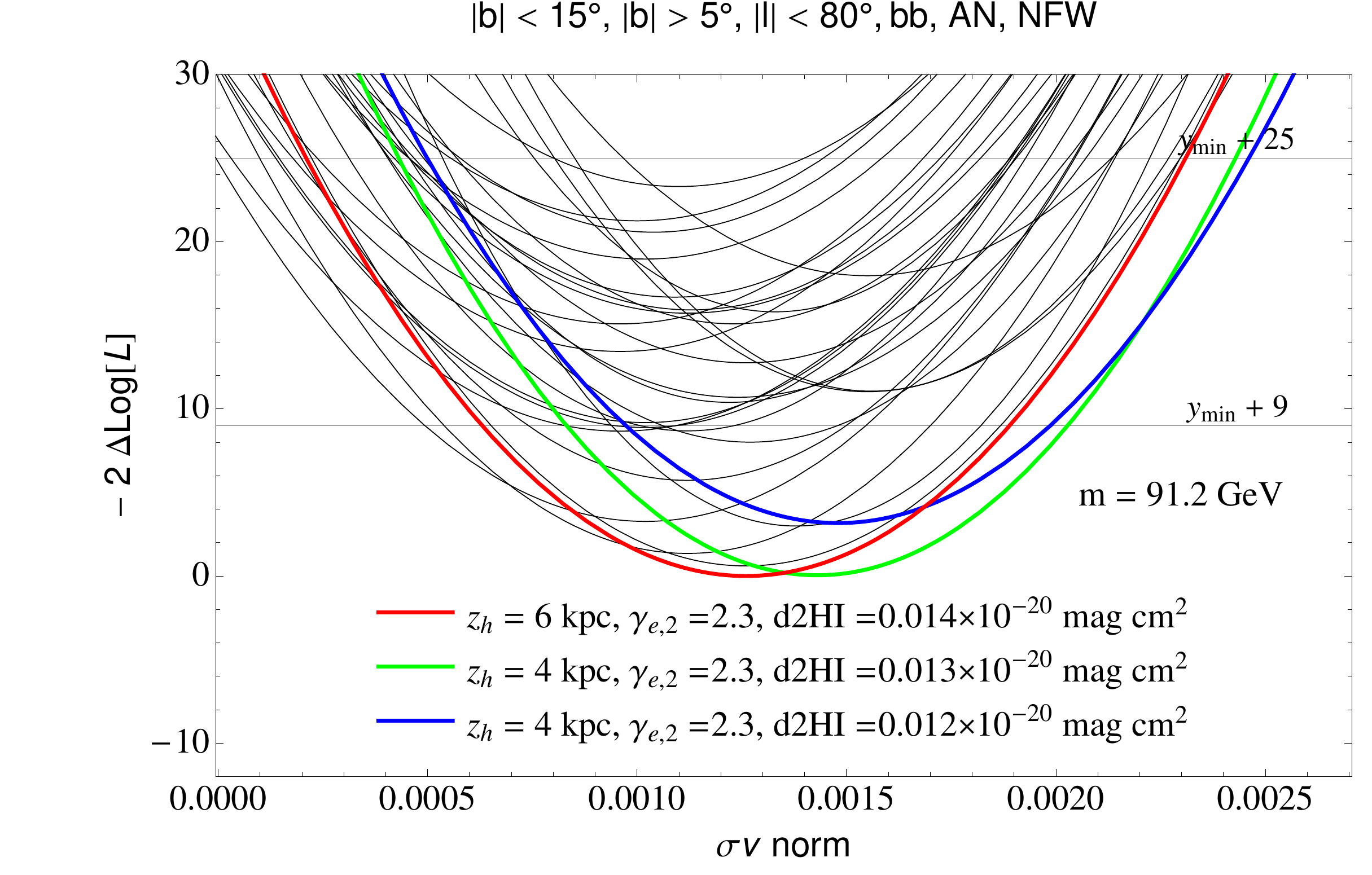}\\
\includegraphics[width=0.45\textwidth]{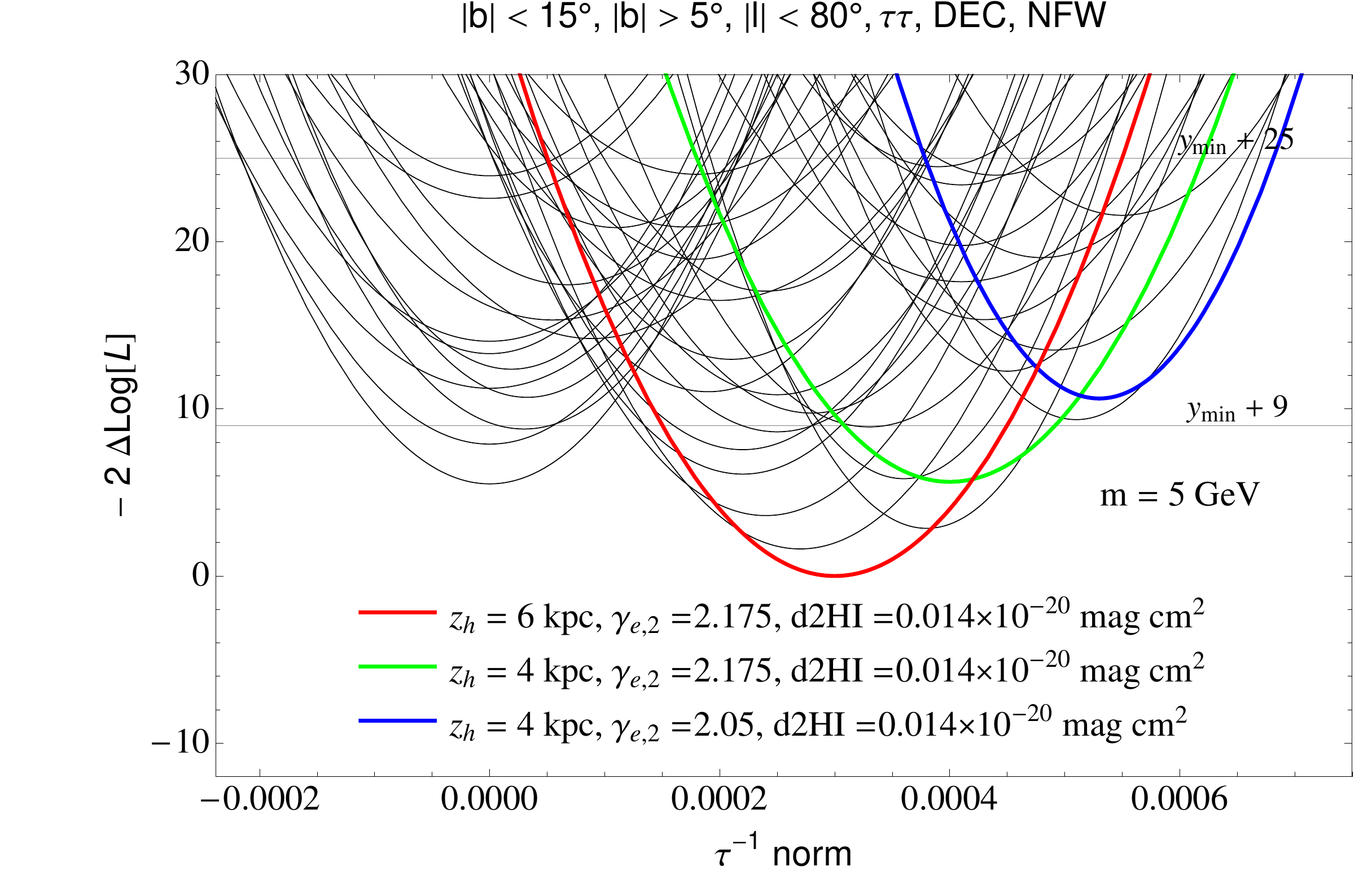}&
\includegraphics[width=0.45\textwidth]{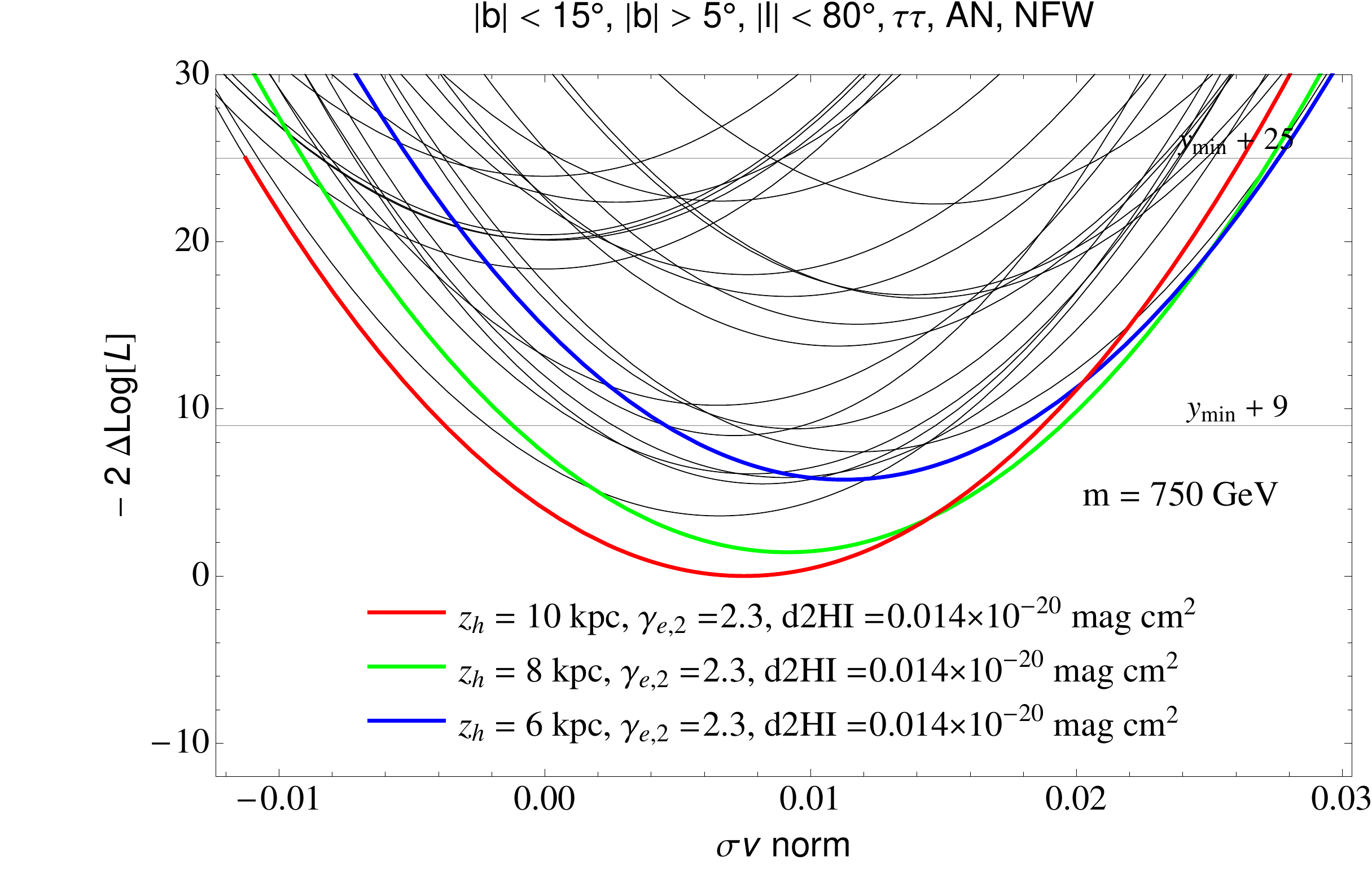}
\end{array}$
\end{center}
\caption{Example of profile likelihood curves for four different DM annihilation/decay scenarios. Each curve refers to a particular model of the background. The envelope of the various curves approximates the global profile likelihood marginalized over the astrophysical uncertainties accounted for in our fitting procedure. The curve corresponding to the model  setting the global minimum, $y_{min}$, is highlighted in red. The $y$ scale is arbitrarily re-shifted so that the minimum value is zero.  The green curve corresponds to the model setting the 3 $\sigma$ upper limit (i.e. the model which is both part of the envelope profile likelihood and intersects the horizontal line located at $+9$). The  upper limit is then effectively given by the $x$ coordinate of the intersection point. The blue curve is similar, but for the 5 $\sigma$ case (and intersects the horizontal line located at $+25$). For these 3 models the corresponding values of $z_h$, $\gamma_{e,2}$, and d2HI are given in  the caption. 
\emph{Panel description:} 10 GeV DM particle decaying (DEC) into $b\bar{b}$ and NFW profile (upper left),  91 GeV DM particle annihilating (AN) into $b\bar{b}$ and NFW profile (upper right),  5 GeV DM particle decaying into $\tau^+\tau^-$ and NFW profile (lower left)  and 750 GeV DM particle annihilating into $\tau^+\tau^-$ and NFW profile (lower right).  
\label{fig:parabolae}}
\end{figure*}

\section{DM limits with modeling of astrophysical background}\label{main}

We derive a second set of upper limits taking into account a model of the astrophysical background. 
As described in sections \ref{diffusemodeling} and \ref{outline}, the approach we use is a combined fit of DM and of a parameterized background model and  we consider the uncertainties in the background model parameters through the profile likelihood method described below.

\subsection{Profile Likelihood and grid scanning} \label{profilelikelihood}

For each DM channel and mass the 
model which describes the LAT data best maximizes the likelihood function which is defined as a product running over all spatial and spectral bins $i$ ,
\begin{equation}
 L_{k}(\theta_{DM})= L_k(\theta_{DM}, \hat{\hat{\vec{\alpha}}})= max_{\vec{\alpha}} \prod_{i} P_{ik}(n_{i};\vec{\alpha},\theta_{DM}),
\label{leq}
\end{equation}
where $P_{ik}$ is the Poisson distribution for observing $n_i$ events in bin $i$ given an expectation value that depends on the parameter set ($\theta_{DM}$, $\vec{\alpha}$). $\theta_{DM}$ is the intensity of the DM component, 
$\vec{\alpha}$ represents the set of parameters which enter the astrophysical diffuse emission model as linear pre-factors to the individual model components (cf. equation \ref{eq:F} below), while $k$ denotes the set of parameters which enter in a non-linear way. Individual \texttt{GALPROP} models have been calculated for a grid of values in the $k$ parameter space.
For each family of models with the same set of non-linear parameters $k$ the \emph{profile likelihood} curve is defined for each $\theta_{DM}$ as the likelihood which is maximal over the possible choices of the parameters $\vec{\alpha}$ for fixed $\theta_{DM}$ (see \cite{2005NIMPA.551..493R} and references therein).  The notation $\hat{\hat{\vec{\alpha}}}$ represents the conditional maximization of the likelihood with respect to these parameters.
The linear part of the fit is performed with \texttt{GaRDiAn}, which for each  fixed value of $\theta_{DM}$ finds the $\vec{\alpha}$ parameters which maximizes the likelihood and the value of the likelihood itself at the maximum\footnote{Technically, instead of maximizing the likelihood, \texttt{GaRDiAn} minimizes  the (negative of) log-likelihood, -$\log L$, using an external minimizer. For our analyses we used \texttt{GaRDiAn} with the \texttt{Minuit} \citep{1975CoPhC..10..343J} minimizer.} (for details about fitting linear parameters see section \ref{sec:linearpar}). 
However, since building the profile likelihood on a grid of  $\theta_{DM}$ values is computationally expensive, we use an alternative approach including 
$\theta_{DM}$ explicitly in the set of parameters fitted by \texttt{GaRDiAn}. In this case \texttt{GaRDiAn} also computes the $\theta_{DM}$ value which maximize the likelihood (the best fit value ${\theta_{DM0}}$) and its $1\sigma$ error estimated from the curvature of the $\log L_k$ around the minimum.  We then approximate the profile likelihood as a Gaussian in $\theta_{DM}$ with mean $\theta_{DM0}$ and width $\sigma_{\theta_{DM0}}$. We have verified that this approximation works extremely well for a subset of cases for which we also explicitly computed the profile likelihood, tabulating it on a grid of $\theta_{DM}$ values. We will thus use this approximation throughout the rest of the analysis.

In this way we end up with a set of $k$  profiles of likelihood $L_k(\theta_{DM})$, one for each combination of the non-linear parameters. The envelope of these curves then approximates the \emph{final} profile likelihood curve, $L(\theta_{DM})$, where all the parameters, linear and non-linear have been included in the profile\footnote{We will sometime use in the following the term \emph{marginalizing}  although, typically, the term applies only within the framework of Bayesian analyses. In our frequentist approach it is called \emph{profiling}.} . Examples of such final profile likelihood curves for specific DM models can be seen in Figure \ref{fig:parabolae}, and will be discussed more in detail in Sec.~\ref{sec:linearpar}.

Limits are calculated from the profile likelihood function by finding the  $\theta_{DM,lim}$ values for which $L(\theta_{DM,lim})/L(\theta_{DM,max})$ is $\exp(-9/2)$ and $\exp(-25/2)$, for $3$ and 5 $\sigma$ C.L. limits, respectively. 
This approximation is exact for Gaussian likelihood functions in one parameter and, due to invariance of the likelihood function under reparameterization, it is most often also  applicable to the non-Gaussian case \citep{2006smep.book.....J}. For the case of handling nuisance parameters, this is not true a priori, but has been shown to give satisfactory properties for a variety of nuisance parameter configurations (e.g. in \cite{2005NIMPA.551..493R,2010PhRvL.104i1302A,2011PhRvL.107x1302A}).  In particular see also the recent search for the Higgs boson at the Large Hadron Collider, where $\mathcal{O}$(100) nuisance parameters need to be taken into account \citep{2012PhLB..710...49A}. We therefore are confident that this approach gives the desired statistical properties, i.e., good coverage and discovery power,  also in our analysis.

\subsection{Free CR Source Distribution and constrained setup limits}

In this section we introduce the first set of linear parameter, i.e. the coefficients defining the CRSDs. The remaining linear parameters will be introduced in the next section.

As noted in section \ref{diffusemodeling}, CRSDs (for example the ones considered in \cite{2012arXiv1202.4039T}) can be modeled from the direct observation of tracers of SNR, and so can be observationally biased. The uncertainty in the distribution of the tracers in the inner Galaxy is therefore large and should be taken into account in the derivation of the DM limits. We therefore fit the CRSD from the gamma-ray data, as described below. 

Due to the linearity of the propagation equation it is possible to combine solutions obtained from different CRSDs. To exploit this feature we define a parametric CRSD as sum of step functions in Galactocentric radius $R$, with each step spanning a disjoint range in $R$:  
\begin{equation} \label{CRSDsteps}
 e,pCRSD(R)= \sum_i c_i^{e,p} \ \theta(R-R_i)\theta(R_{i+1}-R) 
\end{equation}
We choose 7 steps with boundaries: $R_i=$0, 1.0, 3.0, 5.0, 7.0, 9.0, 12.0, 20.0 kpc.  The expected gamma-ray all sky emission for each of the 14 single-step primary $e$ and $p$ distributions are calculated with \texttt{GALPROP}. It is also worth noting that a different \texttt{GALPROP}  run needs to be done for each set of values of the non-linear parameters, since, for a given  $e,p$CRSD the output depends on the entire propagation setup.
For more accurate output, especially in the inner Galaxy, which we are interested in, \texttt{GALPROP}  is run with a 
finer grid in Galactocentric radius $R$ with $dr=0.1$ kpc, compared to the standard grid of $dr=1$ kpc.  
The coefficients $c_i^{e,p}$ are set to unity for the individual \texttt{GALPROP} runs and then fitted from the gamma-ray data as described below.

In order to have conservative and robust limits  we constrain the parameter space defined above by setting $c_1^{e,p}=c_2^{e,p}=0$, i.e. setting to zero the ${e,p}$CRSDs in the inner Galaxy region, within 3 kpc of the Galactic Center. In this way, potential $e$ and $p$ CR sources which would be required in the inner Galaxy  will be  potentially compensated by DM, producing conservative constraints. 
A second important reason to set the inner ${e,p}$CRSDs to zero is the fact that they are strongly degenerate with DM (especially the inner ${e}$CRSD, see Figure \ref{dmmaps}). Besides slight morphological differences, an astrophysical CRE source in the inner Galaxy is hardly distinguishable from a DM source, apart, perhaps, from differences in the energy spectrum. 
To break this degeneracy we would need to use data along the Galactic Plane (within $\pm5^\circ$ in latitude) since these are expected to be the most constraining for the ${e,p}$CRSDs in the inner Galaxy.  However, the Galactic Center region is quite complex and modeling it is beyond the scope of the current paper. 
We therefore defer such a study to follow-up publications.

\begin{table*}[t]
\begin{center}
\begin{tabular}{ | c|c|c| }
\hline
\textbf{ {Non linear Parameters}} & \textbf{  Symbol}   & \textbf{  Grid values}  \\
\hline \hline
index of the injection CRE spectrum & {\bf $\gamma_{e,2}$}   &     1.925, 2.050, 2.175, 2.300, 2.425, 2.550, 2.675, 2.800  \\
 half height of the diffusive halo\footnote{The parameters $D_0$, $\delta$, $v_A$, $\gamma_{p,1}$, $\gamma_{p,1}$, $\rho_{br,p}$ are varied together with $z_h$ as indicated in Table \ref{pdiffusion}. } &  {\bf $z_h$}   &  2, 4, 6, 8, 10, 15 kpc\\
dust to HI ratio & {\,\, d2HI}  \,\,  &  (0.012, 0.013, 0.014, 0.015, 0.016, 0.017)  $\times 10^{-20}$ mag cm$^2$ \\
\hline
\hline
\textbf{ {Linear Parameters}} & \textbf{  Symbol}   & \textbf{  Range of variation}  \\
\hline \hline
 eCRSD and pCRSD coefficients &  {$c^e_i$,$c^p_i$}  &  0,+$\infty$ \\
 local  H$_2$ to CO factor  & {$X_{CO}^{loc}$}  &  0-50 $\times 10^{20}$ cm$^{-2}$  (K km s$^{-1}$)$^{-1}$\\
 IGB normalization in various energy bins & $\alpha_{IGB,m}$  & free \\
 DM normalization &  {$\alpha_{\chi}$}  &  free \\
\hline
\end{tabular}
\caption{Summary table of the parameters varied in the fit. The top part of the table shows the non linear parameters and the grid values at which the
likelihood is computed. The bottom part shows the linear parameters and the range of variation allowed in the fit. The coefficients of the CRSDs
are forced to be positive, except  $c^{e,p}_1$ and $c^{e,p}_2$  which are set to zero. The local $X_{CO}$ ratio is restricted to vary in the range 0-50 $\times 10^{20}$ cm$^{-2}$  (K km s$^{-1}$)$^{-1}$, while $\alpha_{IGB,m}$ and $\alpha_{\chi}$ are left free to assume both  positive and negative values. See the text for more details.}
\label{tab:summarypar}
\end{center}
\end{table*}

\begin{figure*}[t] 
\begin{center}$
\begin{array}{cc}
%\hspace{-2pc}
\includegraphics[width=0.45\textwidth]{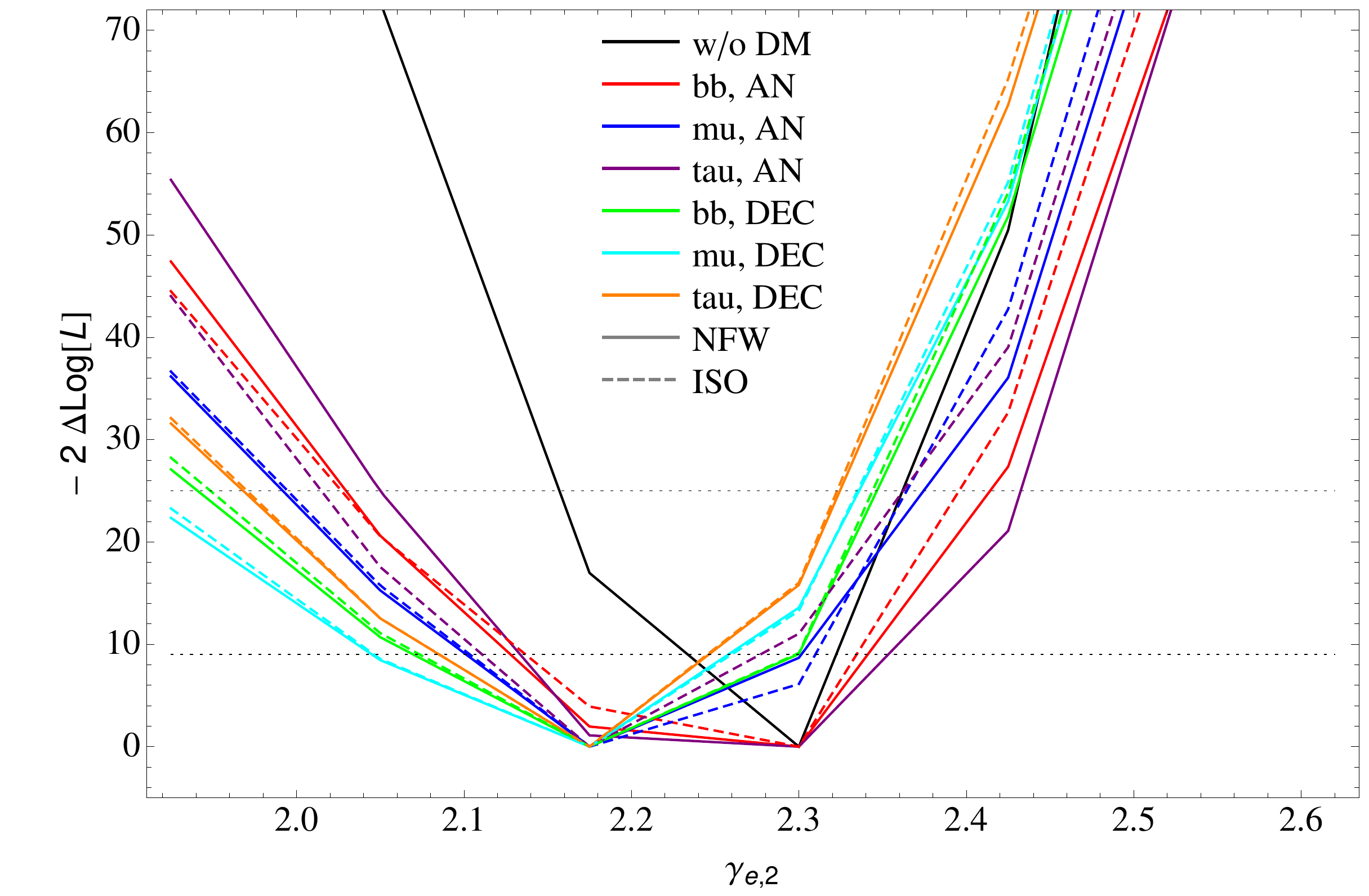}&
\includegraphics[width=0.45\textwidth]{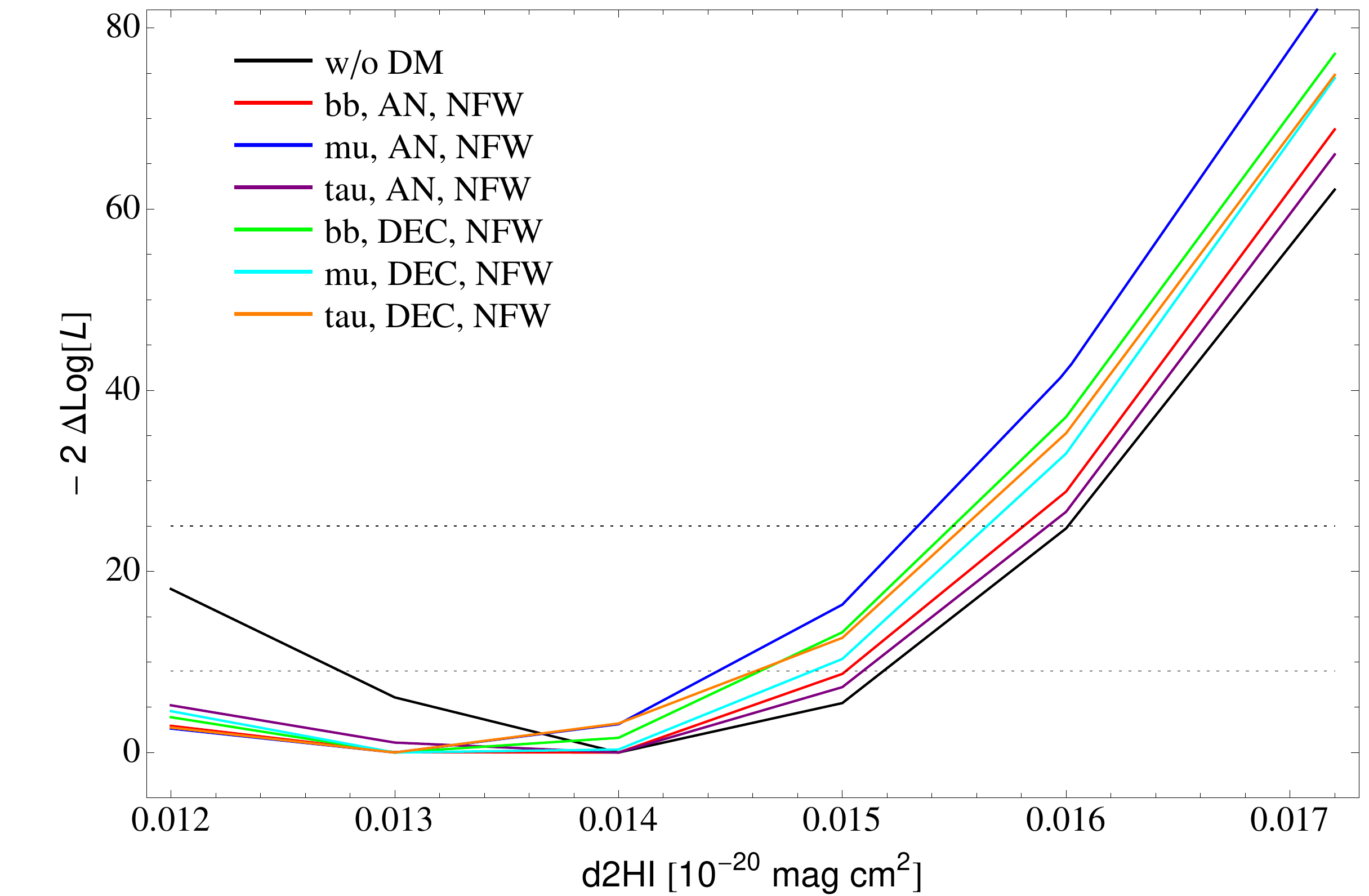}\\
\end{array}$
\end{center}
\begin{center}$
\begin{array}{c}
%\hspace{-2pc}
\includegraphics[width=0.45\textwidth]{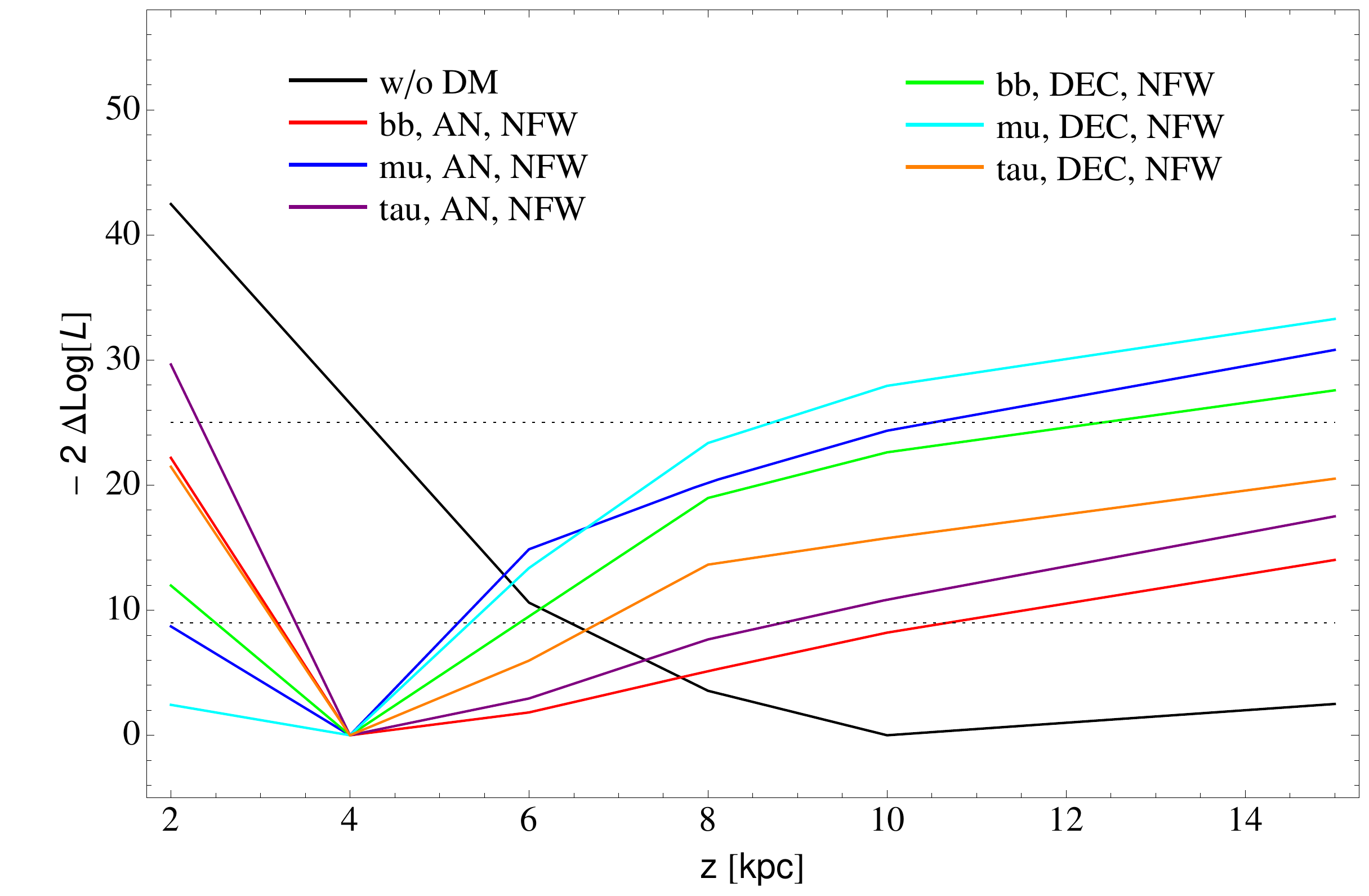}
\end{array}$
\end{center}
\caption{Profile likelihood curves for $z_h$, $\gamma_{e,2}$ and d2HI. The various curves refer to the case of no DM or different DM models (see the legend in the figure, where we mark a dominant decay (DEC) or annihilation (AN) channel and the assumed DM profile). All minima are normalized to the same level. Horizontal dotted lines indicate, as in Figure \ref{fig:parabolae},   a difference in  $-2\Delta{\rm log}L$ from the minimum of 9 (3$\sigma$) and 25 (5$\sigma$).  \label{fig:profiles}}
\end{figure*}

\subsection{Fitting procedure}  \label{sec:linearpar}

In the fit of the expected gamma-ray emission to the {\it Fermi} LAT data we determine the normalizations of the contributions from DM and from each step of the CRSD function defined in Eqn. \ref{CRSDsteps} that best fit the data. To achieve this, we need to split each contribution into several components corresponding to the type of target and physical process responsible for the emission. The emission from $\pi ^0$ decay depends only on the distribution of the CR nuclei sources, while the emission from bremsstrahlung and inverse Compton depends only on the distribution of the CR electron sources (emission from interactions of secondary electrons produced in CR nuclei interactions is negligible above 1 GeV). The gamma-ray emission arising from interactions of CRs with molecular gas traced by CO depends further on the assumed conversion factor $X_{CO}$ between the CO intensity and the column density of the molecular gas. This conversion factor is uncertain and we vary it freely for each annulus. We determine effective $X_{CO}$ factors implicitly in the fit by splitting the calculated expected gamma-ray emission from CR interactions with molecular gas into Galactocentric annuli which are separately normalized. Additionally an isotropic component arising from the extragalactic gamma-ray background and misclassified charged particles needs to be included to fit the {\it Fermi} LAT data. We do not include sources in the fit as we use a mask to filter the 1FGL point sources (cf. Sec \ref{sec:ROI}). To rule out the possibility that some bright sources might leak out of the mask and bias the fit we performed test fits including explicitly the 1FGL point sources as a further template map, finding that the inclusion of the point sources introduces only a negligible change in the results. Equation \ref{eq:F} summarizes how we parametrize the expected gamma-ray emission $I$ in the fit based on the components mentioned above. Each component is calculated using \texttt{GALPROP} and is available as a template map after the \texttt{GALPROP} run.
In summary, the various \texttt{GALPROP}  outputs are combined as:
\begin{eqnarray}\nonumber
     I & = &  \sum_{i}    \bigg\{ c_i^p ~\big( H^i_{\pi^0} +  {\sum_{j}}~X_{\rm CO}^j H_{2 \ \! {\pi^0} }^{ij} \big) +         \\ \nonumber         
       & &    c_i^e ~ \big( H^i_{\rm bremss} +  {\sum_{j}}~X_{\rm CO}^j H_{2\ \! {\rm bremss}}^{ij}  + IC^i \big) \bigg\}  +          \\ \label{fiteq}
       & &   \alpha_\chi~(\chi_\gamma + \chi_{ic}) +  \sum_{m} \alpha_{IGB,m}~IGB^m .
       \label{eq:F}
\end{eqnarray}
The sum over $i$ is the sum over all step-like CRSD functions, the sum over $j$ corresponds to the sum over all Galactocentric annuli (details of the procedure of a placement of the gas in Galactocentric annuli and their boundaries are given in \cite{2012arXiv1202.4039T}). $H$ denotes the gamma-ray emission from atomic and ionized interstellar gas
while $H_2$ the one from molecular hydrogen and $IC$ the Inverse Compton emission.
$\chi_\gamma$ and $ \chi_{ic}$ are the prompt and Inverse Compton (when present) DM contribution and $\alpha_\chi$ the overall DM normalization. 
$IGB^m$ denote the Isotropic Gamma-ray Background (IGB) intensity for each of  the five energy bins over which the index $m$ runs. For better stability of the fit the template for $IGB^m$ is build starting from an IGB with a power law spectrum and normalization as given in \cite{2010PhRvL.104j1101A}. In this way the fit coefficients $\alpha_{IGB,m}$ are typically of order 1.
In all the rest of the expression in Eqn. \ref{eq:F} the energy index $m$ is implicit since we don't allow for the freedom of varying the \texttt{GALPROP}  output from energy bin to energy bin. 
Finally, it should be also noted that in our case, where we mask $\pm 5^{\circ} $  along the plane, the above expression actually simplifies considerably 
 since only the local ring $X_{\rm CO}$ factor enters the sum, since all the other $H_2$ rings do not extend further than 5 degrees from the plane.
Also to be noted is the fact that, since in Eqn. \ref{eq:F} $H_2$ denotes a gamma-ray emission map, the expression has been already intrinsically multiplied by an $X_{\rm CO}$ factor to convert the CO line intensity into an $H_2$ column density. We in fact normalize all the $H_2$ gamma-ray maps using the value  $X_{\rm CO}=1\times 10^{20}$ cm$^{-2}$  (K km s$^{-1}$)$^{-1}$. The $X_{\rm CO}$ in  Eqn. \ref{eq:F}  are thus adimentional ratios with respect to the reference value $1\times 10^{20}$ cm$^{-2}$  (K km s$^{-1}$)$^{-1}$. With a slight abuse of notation we denote them also as $X_{\rm CO}$ factors.

The above expression predicts the expected gamma-ray counts in terms of the  
parameters ($c_i^{e,p}$, $X_{\rm CO}^j$,  $\alpha_{IGB,m}$ and $\alpha_\chi$ for a total of 7+7+1+5+1=21 parameters).
\texttt{GaRDiAn} is used to build the profile likelihood for the intensity of the DM component $\alpha_{\chi}$ by finding the set of parameter values which maximize the likelihood for a given $\alpha_{\chi}$.

The outlined procedure is then repeated for each set of values of the non-linear propagation and injection parameters to obtain the full set of  
profile likelihood curves. We scan over the following three parameters:
the half-height of the diffusive zone $z_h$, the index of the electron injection spectrum $\gamma_{e,2}$ and the dust-to-H~{\sc i} ratio d2HI. 
Specifically, we choose 6 values of $z_h=2,4,6,8,10,15$ kpc, 8 values of $\gamma_{e,2}$ linearly spaced between 1.925 and 2.8, and 6 values of d2HI linearly spaced in the range (0.0120 - 0.0170) $\times 10^{-20}$ mag cm$^2$.  Taking into account the 7 step functions used for the $e,p$CRSDs we scan over a grid of 7$\times$5$\times$8$\times$6=1680 \texttt{GALPROP}  models (or rather \texttt{GALPROP} runs since combinations of the steps are effectively a single \texttt{GALPROP} model.). 

In order to follow more easily the entire fitting procedure we report in Table~\ref{tab:summarypar} a summary of all the parameters employed in our analysis, linear and non-linear, together with their range of variation in the fit or discrete values used in the grid.

Figure \ref{fig:parabolae} shows some examples of the profile likelihoods for selected DM masses and annihilation channels. The limits are set by first finding the absolute minimum  and then looking at  the intersection between   the envelope of the various parabolae  and the  3 and 5 $\sigma$ horizontal lines. An important point to note is that, for each DM model, the global minimum we found lies within the 3(5) $\sigma$ regions of many different models. This is a basic sanity check against a bias in our procedure, as would be suspected if the model giving the minimum was inconsistent with the bulk of the other models considered. This point is further illustrated in Figure~\ref{fig:profiles}, where the  profile likelihoods for the three nonlinear parameters, $z_h$, $\gamma_{e,2}$ and d2HI, are shown. 
To ease reading of the figure the profiling is actually performed with further grouping DM models with different DM masses, but keeping the different DM channels, DM profiles and the annihilation/decay cases separately.  The curve for the fit without DM is also shown for comparison. Each resulting curve has been further rescaled to a common minimum, since we are interested in showing that several models are within $-2\Delta{\rm log}L\leq 25$ around the minimum for each DM fit.
The $\gamma_{e,2}$ profile, for example, indicates that all models with $\gamma_{e,2}$ from 1.9 to 2.4 are within $-2\Delta{\rm log}L\leq 25$ around the minimum illustrating that the sampling around each of the minima for the six DM models is dense. Similarly, the d2HI profile indicates that all models with d2HI in the range  (0.120 - 0.160) $\times 10^{-20}$ mag cm$^2$ are within $5 \sigma$ from the minima for each of the six DM models. 
Finally the $z_h$ profile indicates that basically all the considered values of $z_h$ are close to the absolute minima. This last result is not surprising since, within our low-latitude ROI, we have little sensitivity to different $z_h$ and basically all of them fit equally well.
There is some tendency to favor higher values of $z_h$ when DM is not included in the fit, while with DM the trend is inverted, although the feature is not extremely significant it is potentially very interesting.

As explained in Sec.~\ref{outline}, in our analysis the DM parameter is the one of prime interest and we
thus treat the parameters (linear and non-linear) of the diffuse emission as  \emph{nuisance parameters}
which we include to take into account degeneracies with DM (i.e to marginalize over them) and establish more robust limits.
This is a reasonable assumption since, for example, most of the IC emission and thus the sensitivity to $\gamma_{e,2}$ comes from data within $5^\circ$ of the Galactic Plane, which are not considered here.  Similarly, to be sensitive to $z_h$, high Galactic latitude data should be included. These plots thus should be regarded only as indicative of the achievable constraints, while a careful analysis will be deferred to later publications.

An issue that we are not addressing here is whether DM is required or not in the fit,  and, in the former case, finding the best model among different DM models.  
Since we have seen that systematic uncertainties related to the limitations in modeling astrophysical contributions to the Galactic diffuse emission are comparable in size to any DM signal we fit, we have focussed on setting constraints on potential DM contributions to the Galactic diffuse emission. The systematic uncertainties in the Galactic diffuse emission modeling likely could be reduced by including the Galactic Plane/Galactic Center data in the analysis. Furthermore, a realistic study of the problem would require also considering other possible components which might be present in the Halo, like contributions from a population of unresolved pulsars or the emission from the Bubbles/Lobes. We defer this analisys to a subsequent study.

\section{Results}\label{results}

\begin{figure*}[!tp] 
\begin{center}$
\begin{array}{cc}
\includegraphics[width=0.5\textwidth]{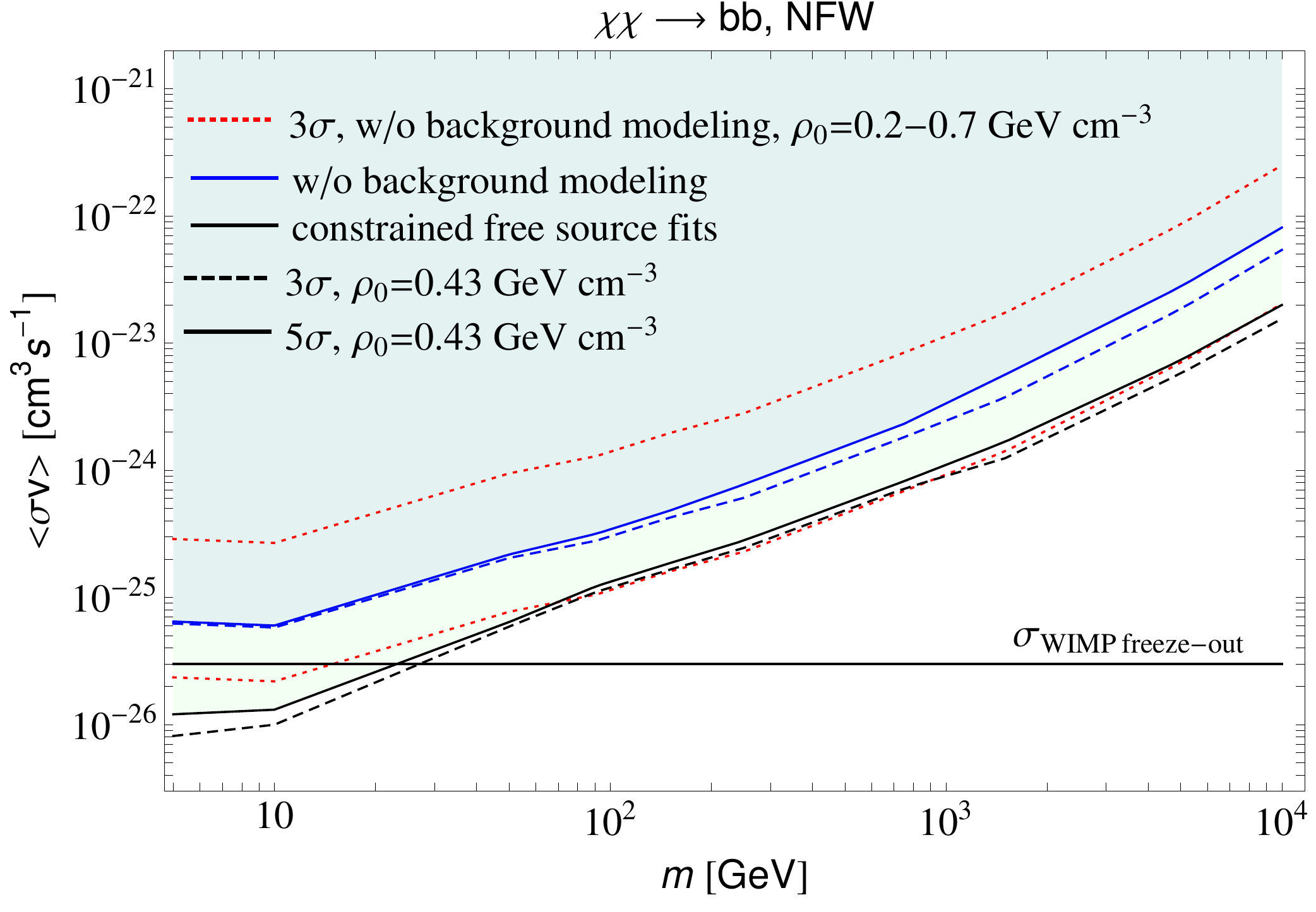}&
\includegraphics[width=0.5\textwidth]{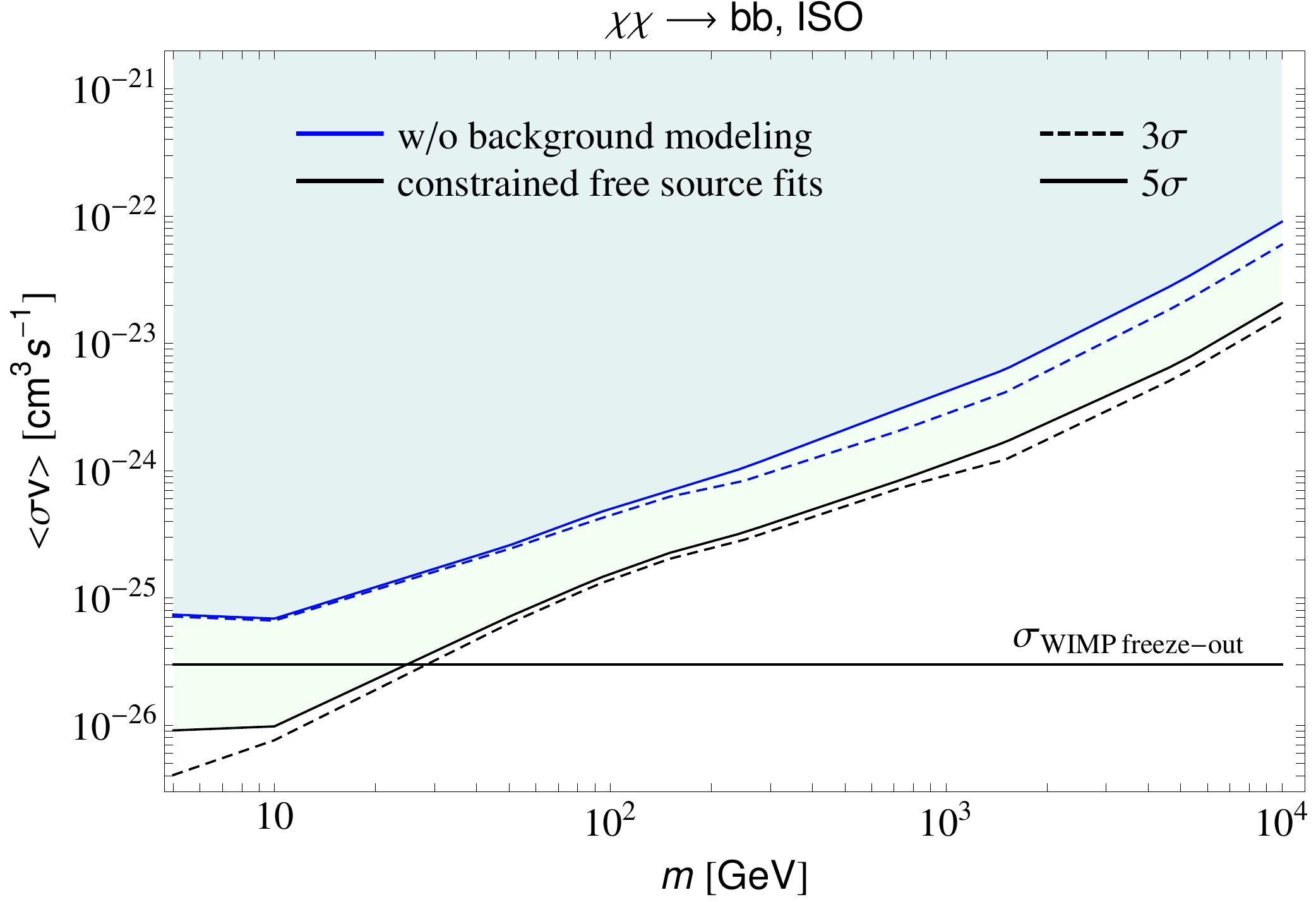}
\end{array}$
\end{center}
%
%\vspace{-1.0cm}
%
\begin{center}$
\begin{array}{cc}
\includegraphics[width=0.5\textwidth]{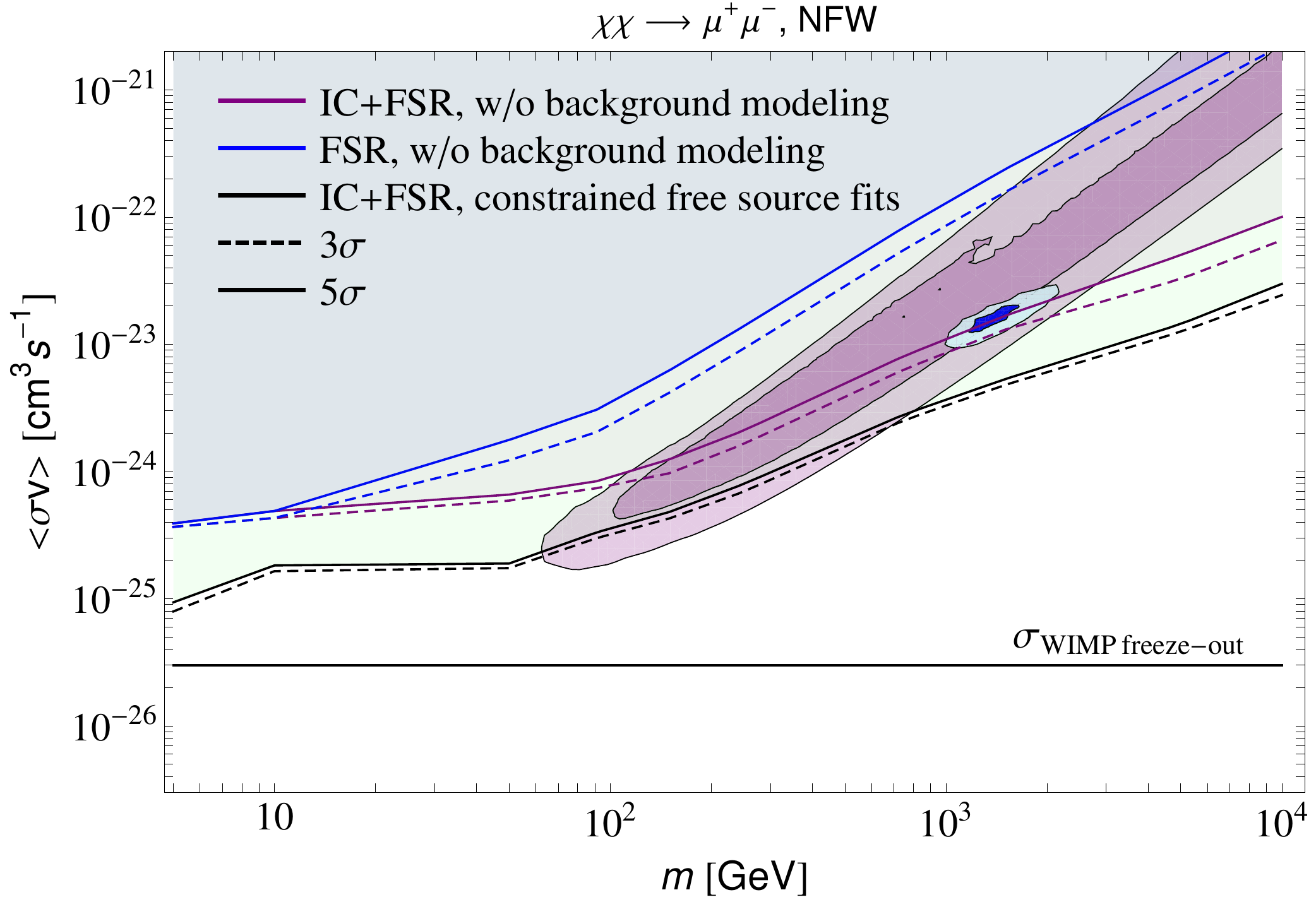}&
\includegraphics[width=0.5\textwidth]{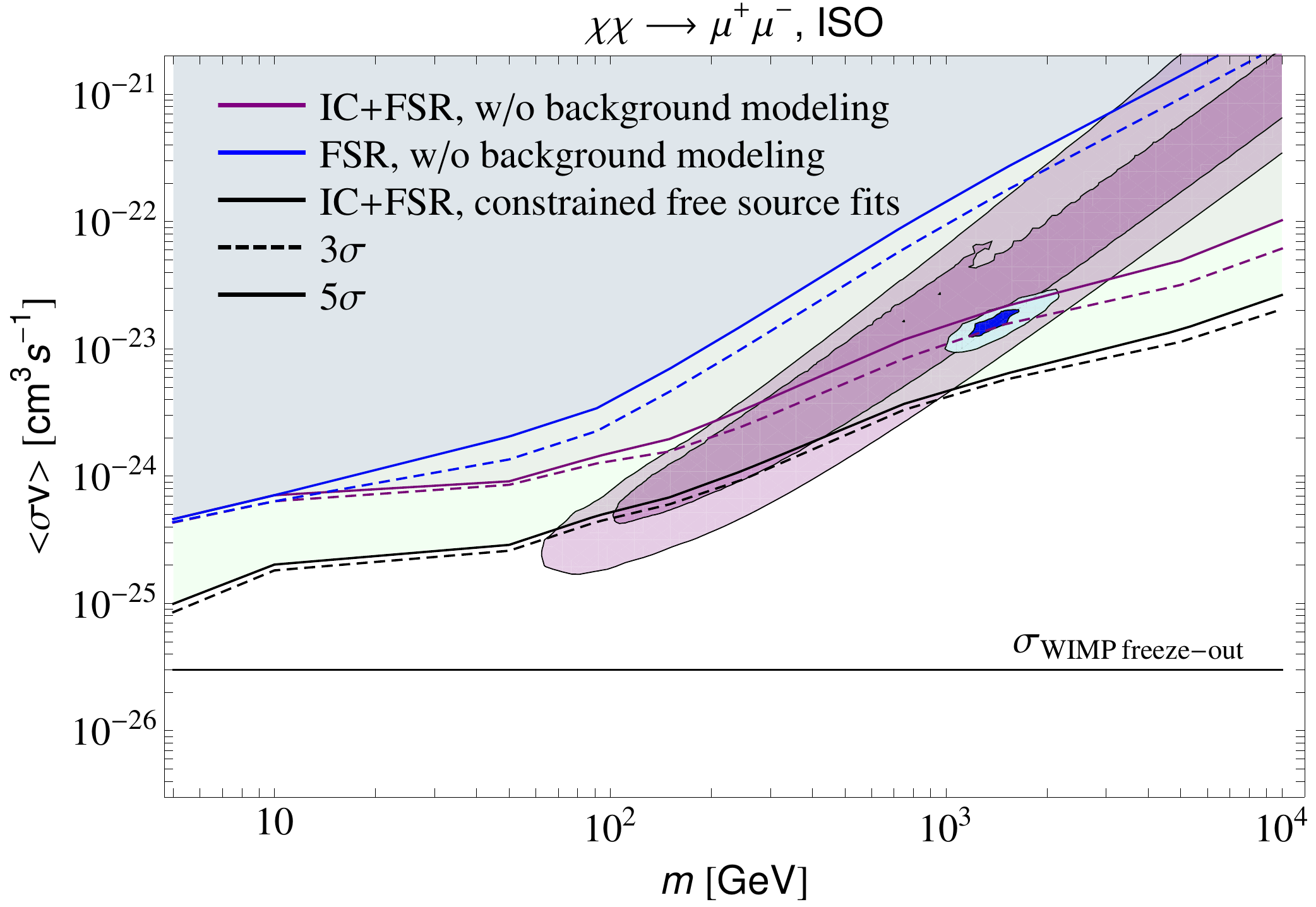}
\end{array}$
\end{center}
\begin{center}$
\begin{array}{cc}
\includegraphics[width=0.5\textwidth]{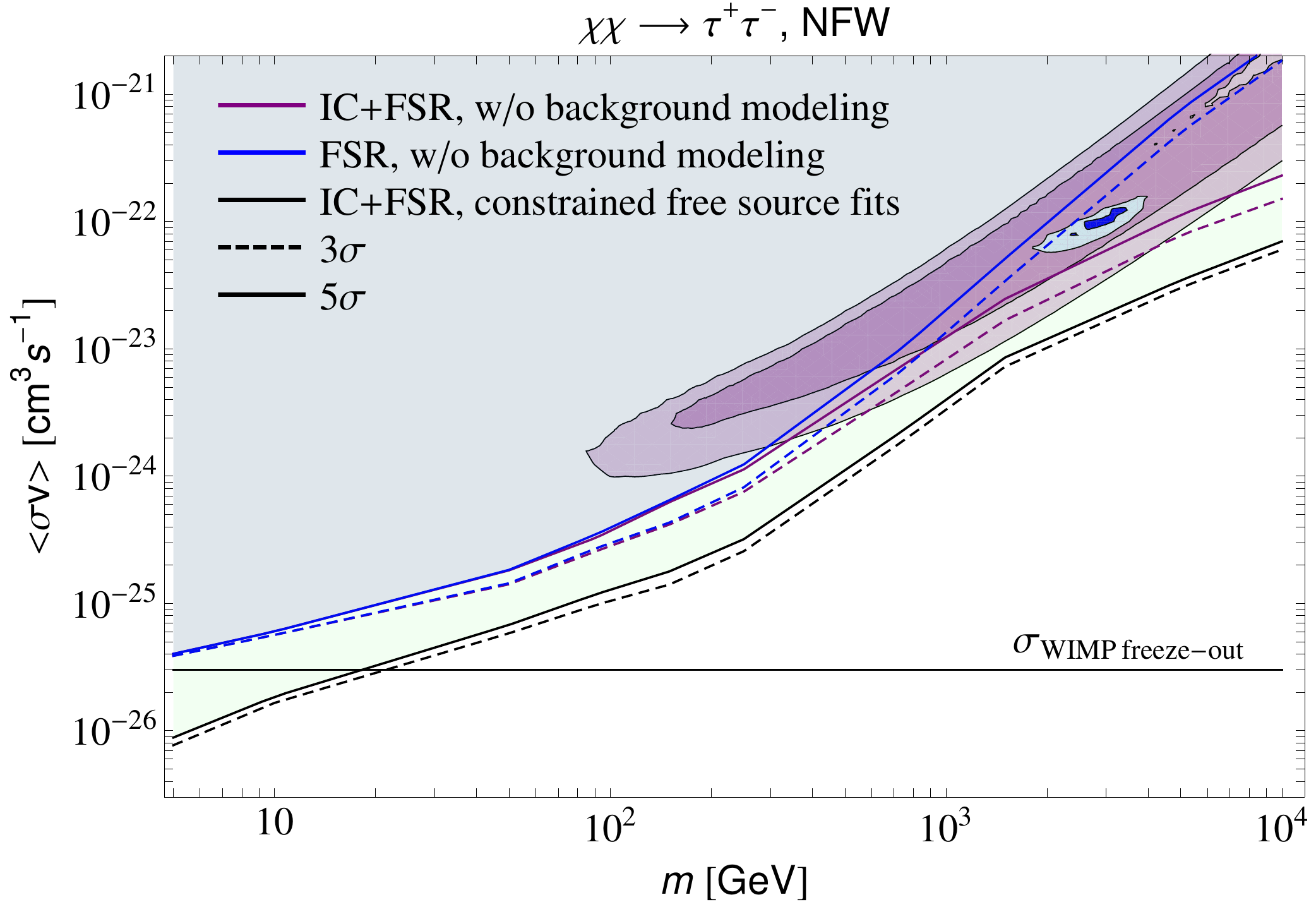}&
\includegraphics[width=0.5\textwidth]{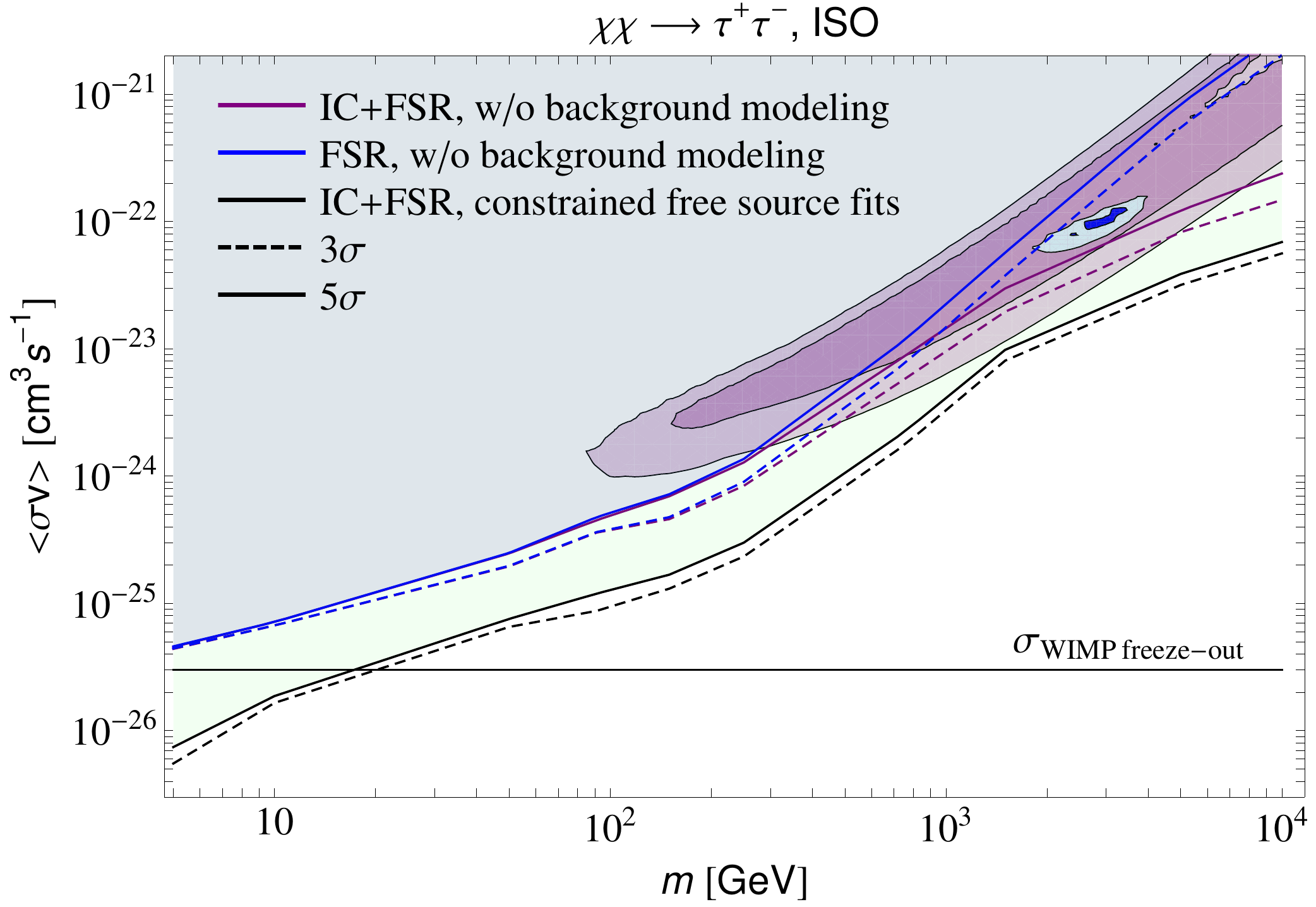}
\end{array}$
\end{center}
\vspace{-0.6cm}
\caption{Upper limits on the velocity averaged DM annihilation cross-section including a model of the astrophysical background compared with the limits obtained with no modeling of the background. Upper panel: Limits on models in which DM annihilates into $b{\bar b}$, for a DM distribution given by the NFW distribution (left) and isothermal distribution (right). In the left panel we also add an uncertainty band (red dotted lines) in the $3\sigma$ no-background limits  which would result from varying the local DM density $\rho_0$ in the range 0.2-0.7 GeV cm$^{-3}$. A similar band, not shown in the plot for clarity, would be present for the limits including a model of the astrophysical background   (see discussion in the text). {\it The horizontal line marks the thermal decoupling cross section expected for a generic WIMP candidate.} Middle panel: Upper limits for DM annihilation to $\mu ^+ \mu^-$. Lower panel: The same, for DM annihilation to $\tau ^+ \tau^-$.  The region excluded by the analysis with no model of the astrophysical background is indicated in light blue, while  the additional region excluded by the analysis with a modeling of the background is indicated in light green.   The regions of parameter space which provide a good fit to PAMELA \cite{2009Natur.458..607A} (purple) and {\it Fermi} LAT \cite{2009PhRvL.102r1101A} (blue) CR electron and positron data are shown, as derived in \cite{2010NuPhB.840..284C} and are scaled by a factor of 0.5, to account for different assumptions on the local DM density (see text for more details).\label{fig:fixedsourcelimits} }
\end{figure*}

Despite the various conservative choices described above, the resulting limits are quite stringent. Upper limits on the velocity averaged annihilation cross section into various channels are shown in Figure~\ref{fig:fixedsourcelimits}, for an NFW and isothermal
profile of the DM halo.  The limits obtained  without modeling of the astrophysical background are compatible with the result of similar analyses presented in  \cite{2010JCAP...03..014P,2010NuPhB.840..284C}. Limits with model of the background, instead, are significantly improved with respect to the above ones.
They are competitive with respect to the limits from LAT 
searches for a signal from DM annihilation/decay in dwarf galaxies  \citep{2011PhRvL.107x1302A} and Galaxy clusters \citep{2010JCAP...05..025A}. 

In particular, as shown in Figure~\ref{fig:fixedsourcelimits} for masses around 20 GeV the thermal relic value of the annihilation cross section is reached,  both for the $b\bar{b}$  and $\tau^+\tau^-$ channels.  
The limits are also improved over the ones derived in  the analysis of dwarf galaxies \citep{2011PhRvL.107x1302A} which did not consider the inverse Compton emission (in dwarfs this component is quite uncertain) 
and also improved over constraints imposed on DM annihilations from the absence of a measurable effect on CMB anisotropies \citep{2011PhRvD..84b7302G}.

\begin{figure*}[!tph] 
\begin{center}$
\begin{array}{cc}
\includegraphics[width=0.5\textwidth]{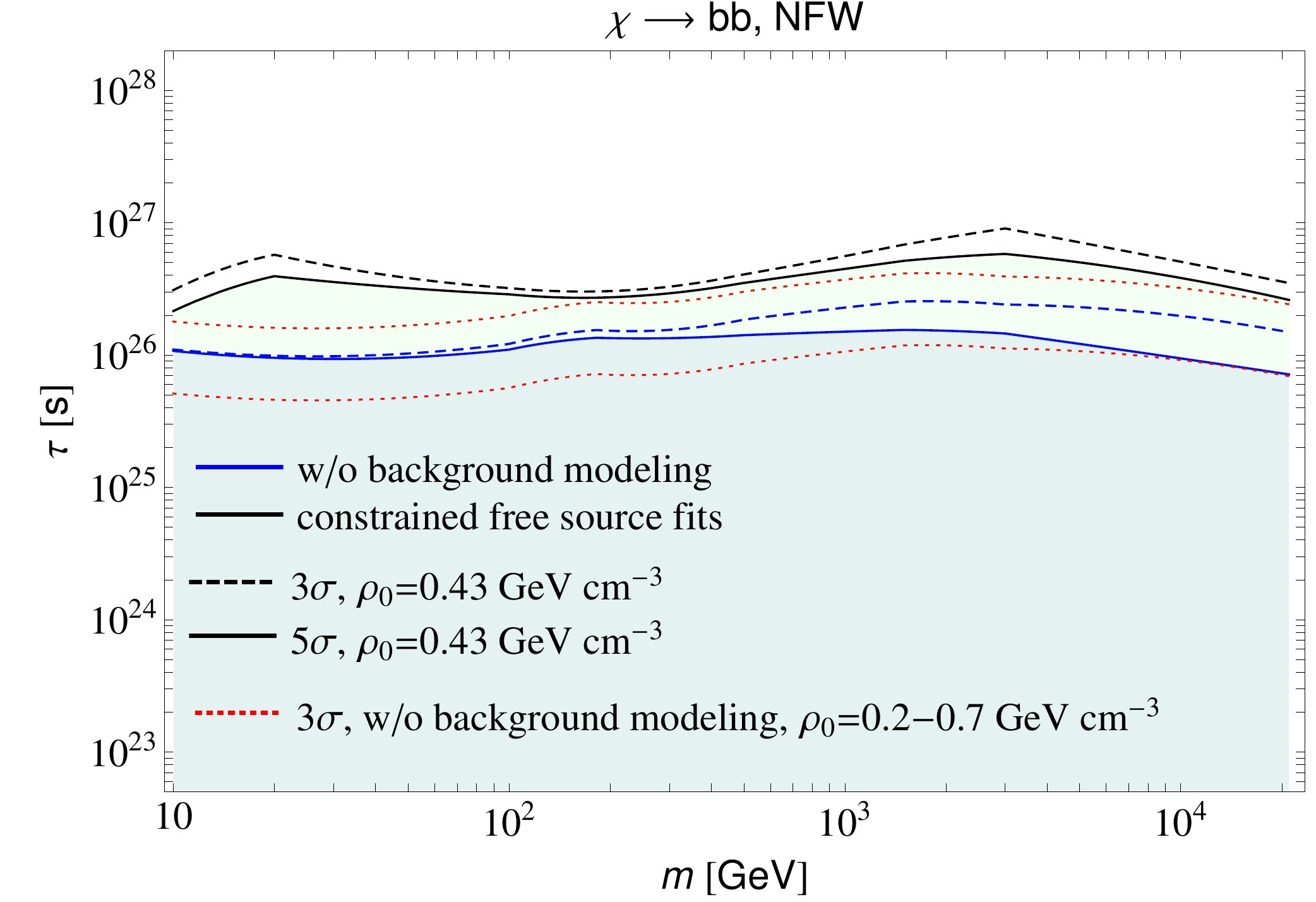}&
\includegraphics[width=0.5\textwidth]{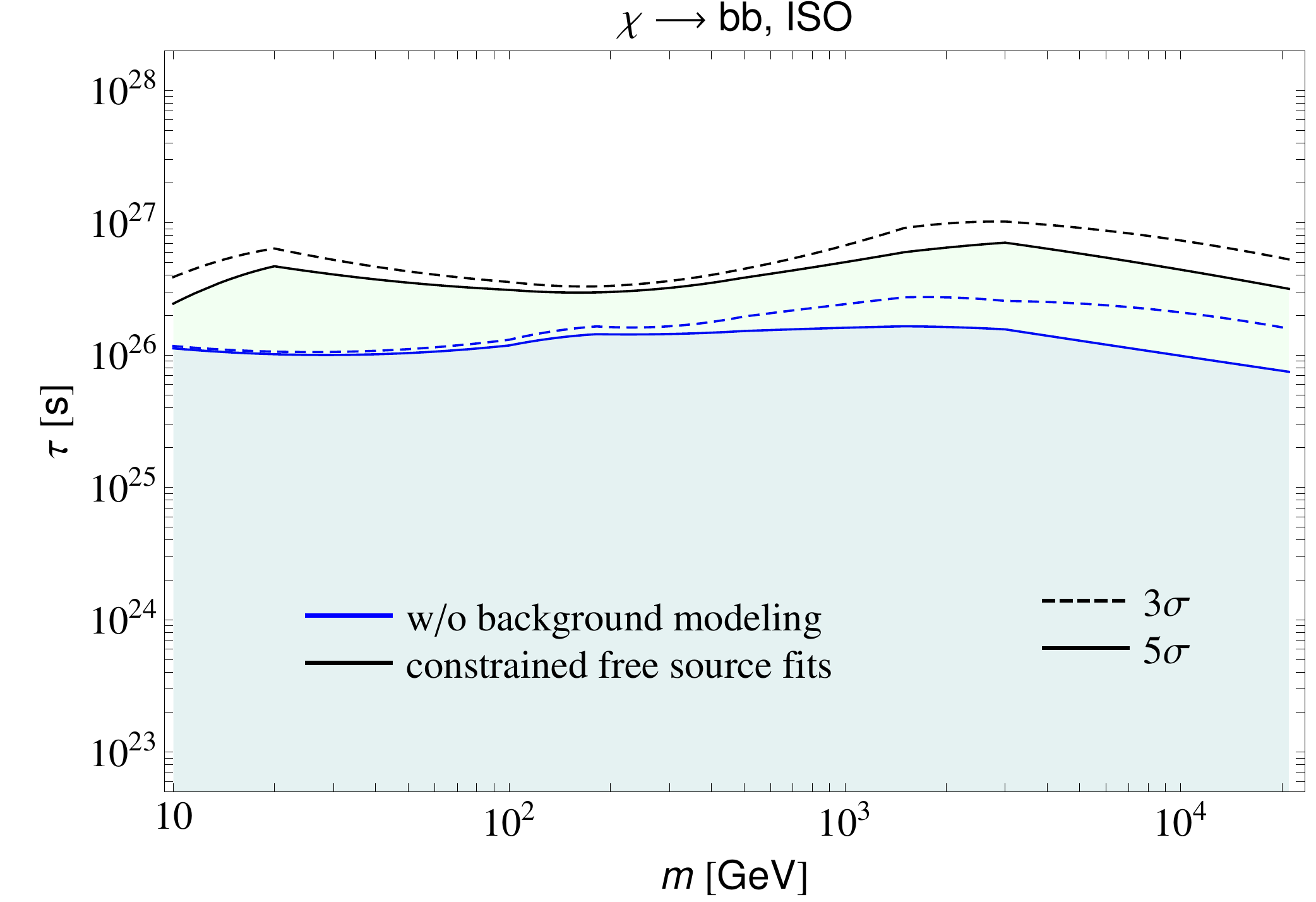}
\end{array}$
\end{center}
\vspace{-0.5cm}
\begin{center}$
\begin{array}{cc}
\includegraphics[width=0.5\textwidth]{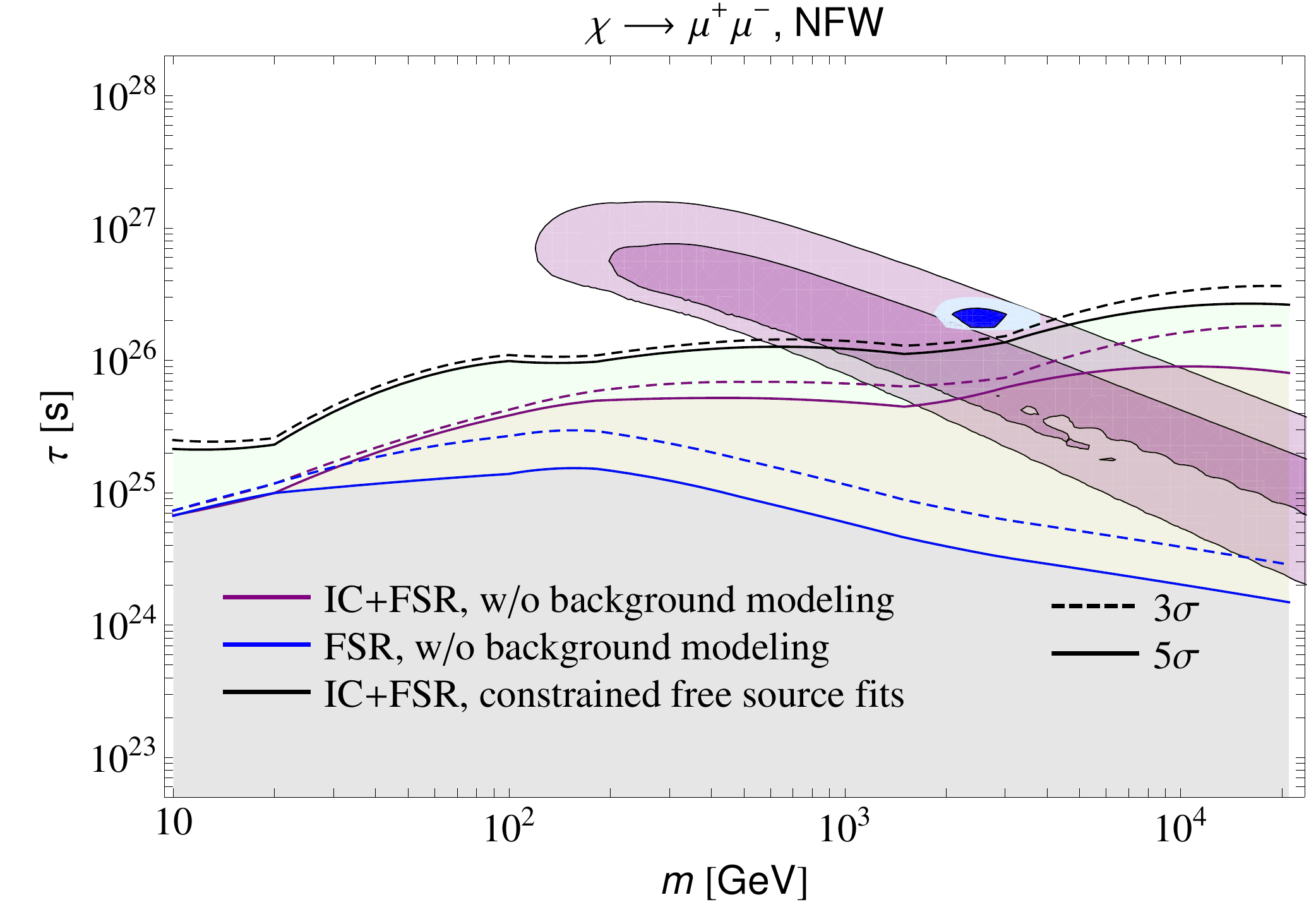}&
\includegraphics[width=0.5\textwidth]{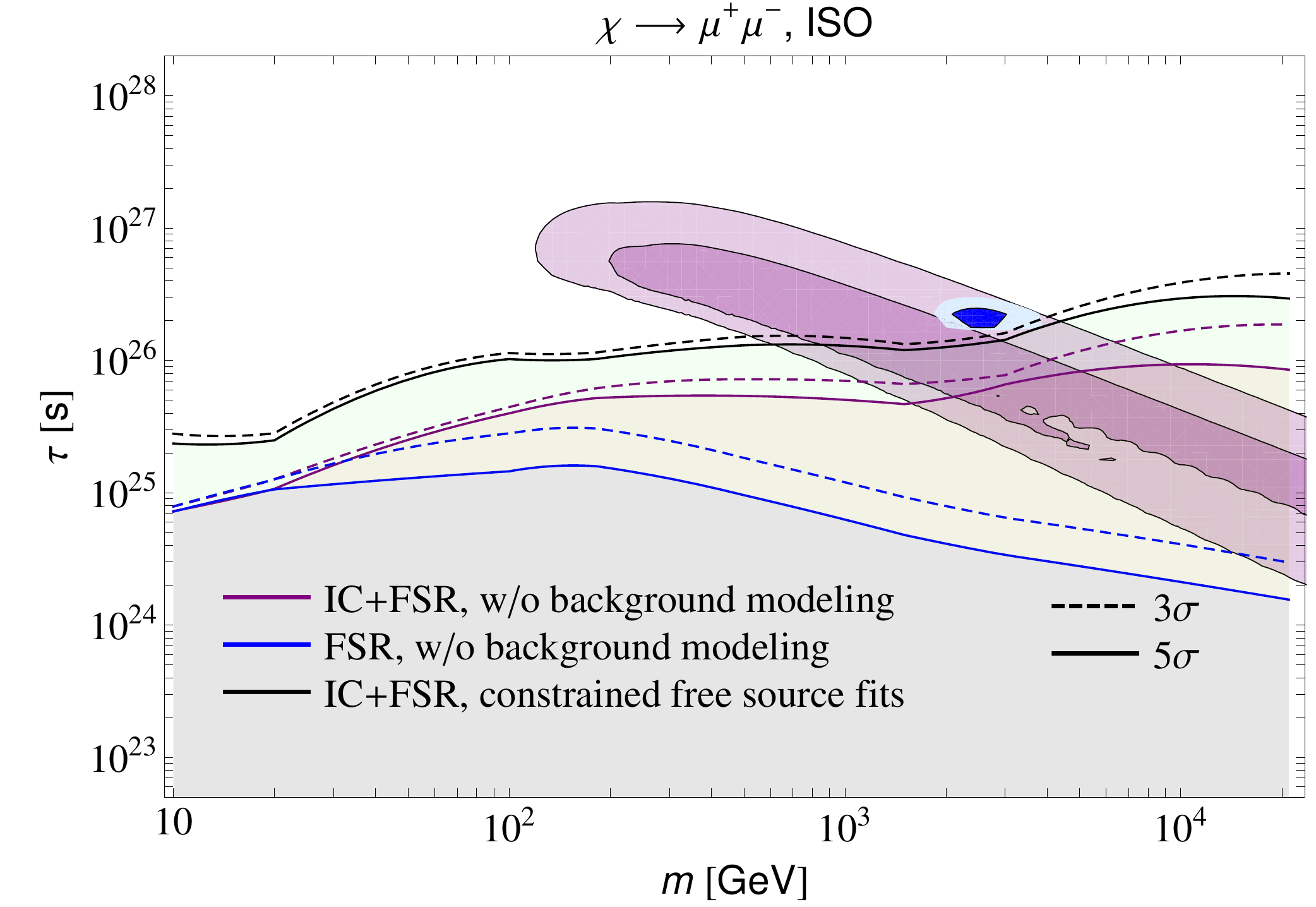}
\end{array}$
\end{center}
\begin{center}$
\begin{array}{cc}
\includegraphics[width=0.5\textwidth]{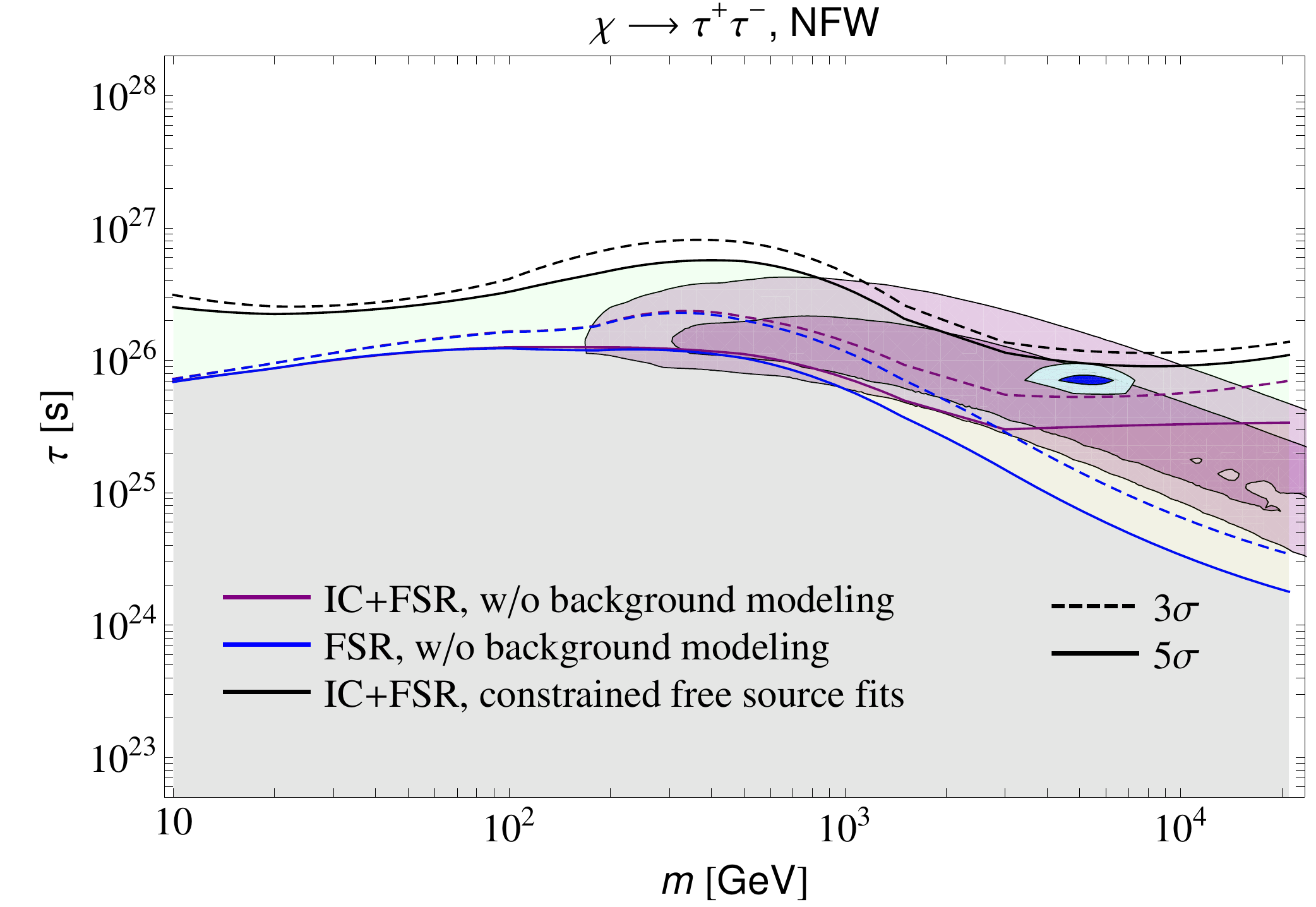}&
\includegraphics[width=0.5\textwidth]{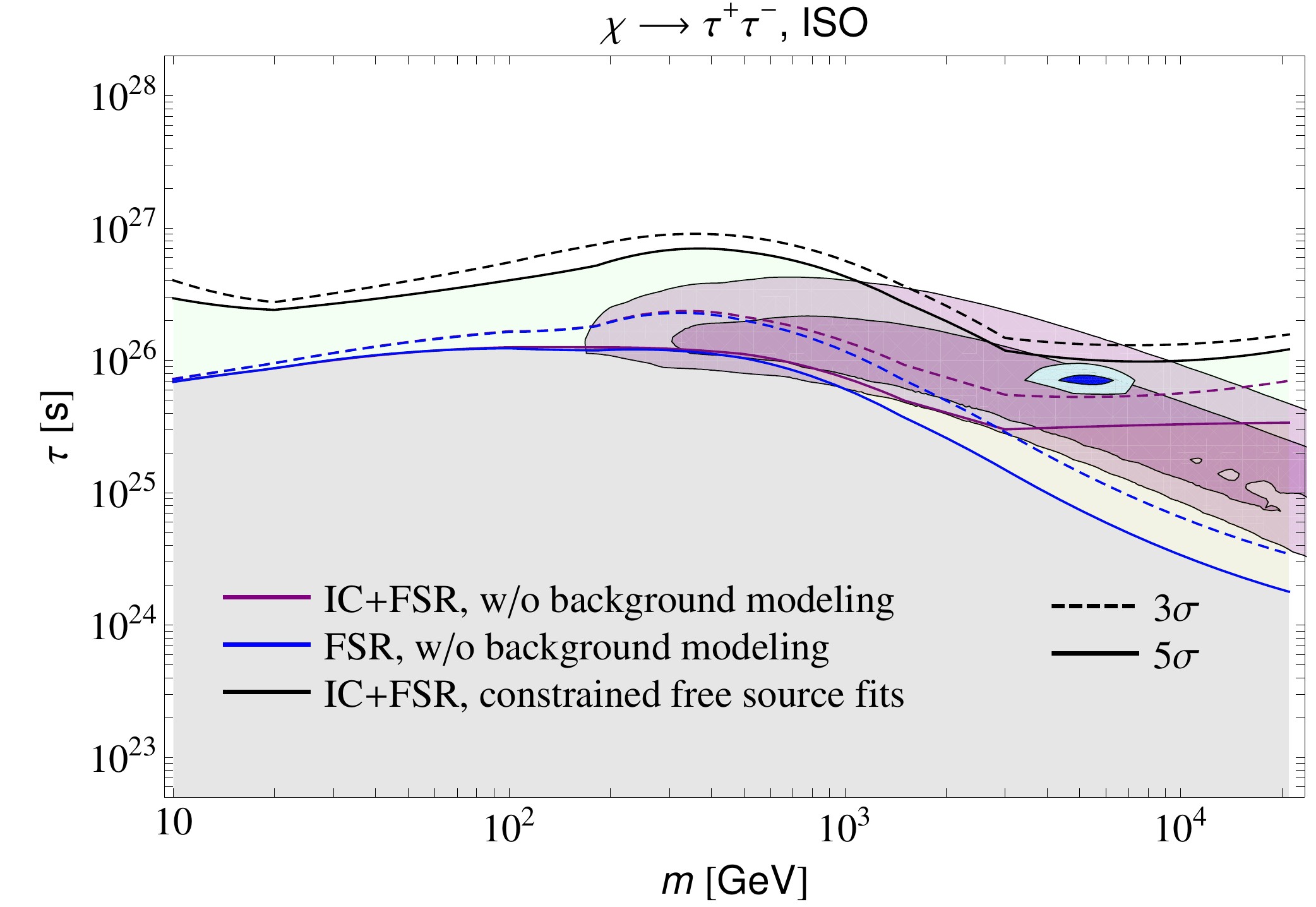}
\end{array}$
\end{center}
\vspace{-0.6cm}
\caption{Lower limits on the lifetime of decaying DM. The panel structure is the same as in Figure \ref{fig:fixedsourcelimits}.  In the top left panel we also add an uncertainty band (red dotted lines) in the $3\sigma$ no-background limits  which would result from varying the local DM density $\rho_0$ in the range 0.2-0.7 GeV cm$^{-3}$. A similar band, not shown in the plot for clarity, would be present for the limits including a model of the astrophysical background   (see discussion in the text). The regions of parameter space which provide a good fit to PAMELA \cite{2009Natur.458..607A} (purple) and {\it Fermi} LAT \cite{2009PhRvL.102r1101A} (blue) CR electron and positron data are shown as derived in \cite{2010NuPhB.840..284C} and are scaled by a factor of 1.4, to account for different assumptions on the local DM density (see text for more details). \label{fig:decay}}
\end{figure*}

A limitation of our constraints is their dependence on poorly determined properties of the Milky Way DM Halo, in particular on $\rho_0$, from which the normalization of the DM signal, and thus the limits, depends quadratically in the annihilation case and linearly in the decay case.
We use the recent determination $\rho_0=0.43$ GeV cm$^{-3}$ from \cite{2010A&A...523A..83S}, which has, however, a large uncertainty, with a
typical associated error bar of $\pm$0.1 GeV cm$^{-3}$ and a possible spread up to 0.2-0.7 GeV cm$^{-3}$ \citep{2010A&A...523A..83S,2011JCAP...03..051C}. Whether the limits will worsen, or improve,   thus awaits a better determination of $\rho_0$. 
To show the effect of the  $\rho_0$ uncertainty on the limits we plot, for illustration, in the top left panel of Figs.\ref{fig:fixedsourcelimits}-\ref{fig:decay}
the uncertainty band (red dotted lines) in the $3\sigma$ no-background limits  which would result from varying the local DM density $\rho_0$ in the range 0.2-0.7 GeV cm$^{-3}$ (we conservatively take here the larger scatter to show the maximal impact of the uncertainty of $\rho_0$). A similar band, not shown in the plot for clarity, would be present for the limits including a model of the astrophysical background. The band is likely a generous estimate of the uncertainty since the variation of $\rho_0$ is typically correlated with other properties of the DM Halo, such as the density profile and the distance of the solar system from the Galactic center $R_S$. 
The uncertainty band shown should just be considered an illustration, while a  detailed study would be required 
to address the actual uncertainty,
which is beyond the scope of this work.

In Figure  \ref{fig:fixedsourcelimits} we also show the regions of the parameter space derived in \cite{2010NuPhB.840..284C} from a  DM fit to the {\it Fermi} LAT electron/positron data and PAMELA positron fraction data. Contours are shown at  95\% and 99.999\% CL. 
These regions are rescaled down of a factor $(0.43/0.3)^2\sim 2$ to take into account the different local DM density $\rho_0$ used in the two works (for the same reason the regions in the decay case are rescaled up by a factor of $(0.43/0.3)\sim 1.4$).
We must also take into account that different energy losses of local CREs are used here when compared to \cite{2010NuPhB.840..284C}. 
For  CREs with $E \geq1~$ TeV diffusion can be approximately neglected
so that, from the diffusion loss equation,  the steady state CRE flux  scales approximately linearly with the energy loss time scale $\tau$ (see \cite{2009PhRvL.103c1103B,2009PhRvD..79b3518B}).  
The PAMELA/{\it Fermi} regions, instead, scale as the inverse of the local CRE flux.
Beyond the different local magnetic field and ISRF, a major difference with the analysis
in \cite{2010NuPhB.840..284C} is that they  neglect Klein-Nishina attenuation effects in the Inverse Compton cross section, which are instead taken into account in our  \texttt{GALPROP} calculations. 
These effects make the energy losses of CREs with $E\geq1$ TeV almost negligible on the optical part of the ISRF, and slightly decrease the energy losses
due the infrared part. CMB losses are instead unaffected. As result, the energy loss time scale, and thus the CRE flux, increases (the PAMELA/{\it Fermi} regions  will be pushed downward).
On the other hand, we use a local magnetic field of $5 \,\mu$G as opposed to the $3\,\mu$G used in  \cite{2010NuPhB.840..284C}, and this has the contrary effect, decreasing the energy losses time scale (which scales as $B^{-2}$) and thus pushing up again the PAMELA/{\it Fermi} confidence regions.
Overall, to correctly derive the positions of the PAMELA/{\it Fermi} confidence regions, the uncertainties in the local ISRF and magnetic field should be taken into account and marginalized away, which we leave for a follow-up work. 
Here, from the above considerations, we estimate that the location of the PAMELA/{\it Fermi} regions has a further uncertainty of a factor $\sim2$, so that they can possibly touch the exclusion limits.
Thus,  we cannot robustly rule out the DM annihilation interpretation of the PAMELA/{\it Fermi} CRs anomalies, although this interpretation is challenged.
Finally, we note that the PAMELA region below $\sim$200 GeV is now disfavored by the new positron measurements with the LAT \cite{2012PhRvL.108a1103A} which indicate that the positron fraction continues to rise to this energy.

It should be noted that the above conclusions are not affected by the uncertainty in $\rho_0$ since both the derived constraints and the region of parameter space compatible with the DM interpretation of the CR anomalies scales in the same way with $\rho_0$. The same is true also for the constrains on decaying DM.

Constraints for the case of decaying DM are shown in Figure  \ref{fig:decay}. The interpretation of the PAMELA/{\it Fermi} CR features in terms of decaying DM is not ruled out in this analysis. The limits are stronger  than the ones derived in  similar analyses performed without background modeling \citep{2010JCAP...03..014P,2010NuPhB.840..284C} and slightly improved over the ones derived from observation of Galaxy clusters \citep{2012JCAP...01..042H}. 
They are comparable to the limits derived from the comparison with the IGB \citep{2010NuPhB.840..284C,2012JCAP...01..042H}.

\begin{figure*}[t] 
\begin{center}$
\begin{array}{cc}
\includegraphics[width=0.45\textwidth]{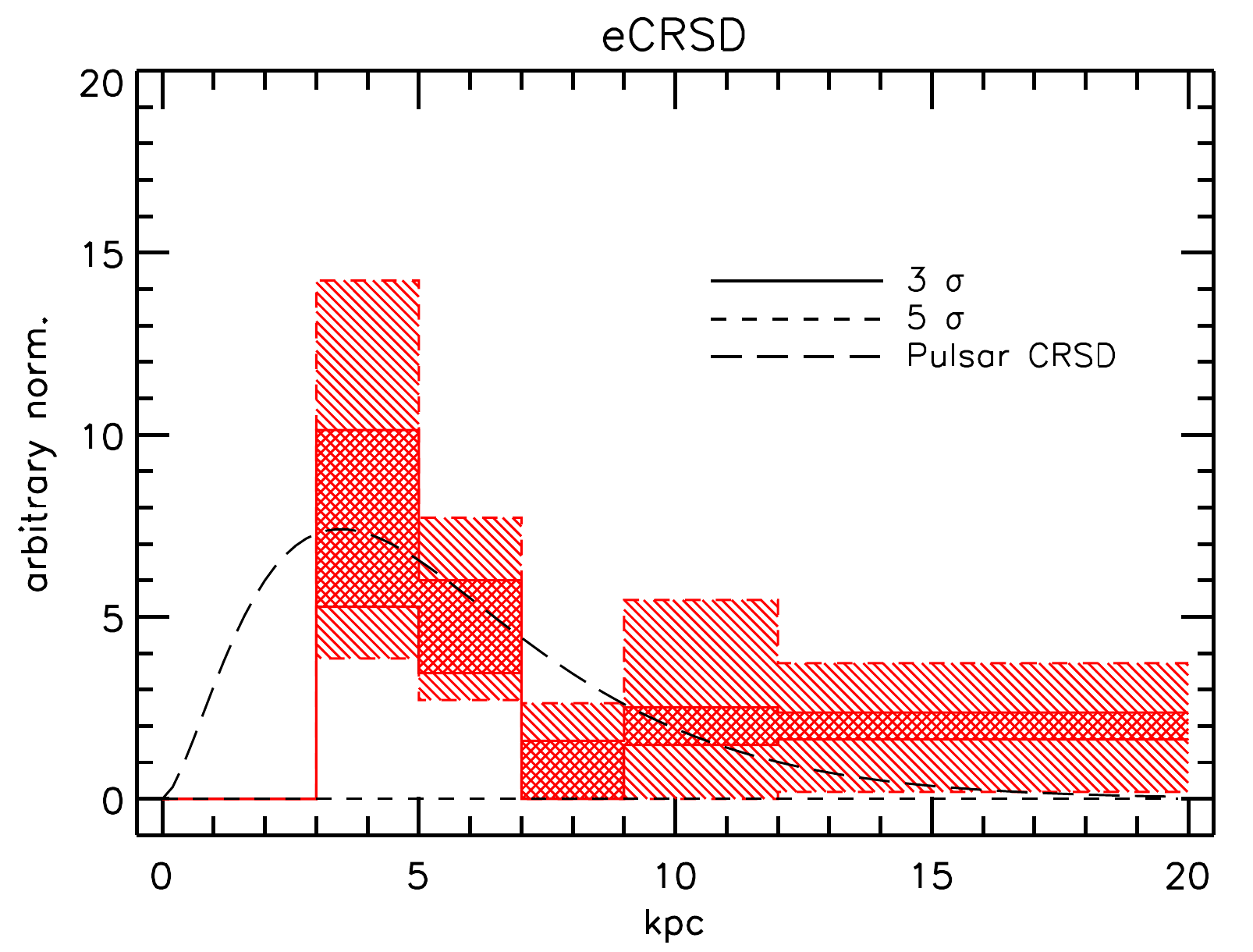}&
\includegraphics[width=0.45\textwidth]{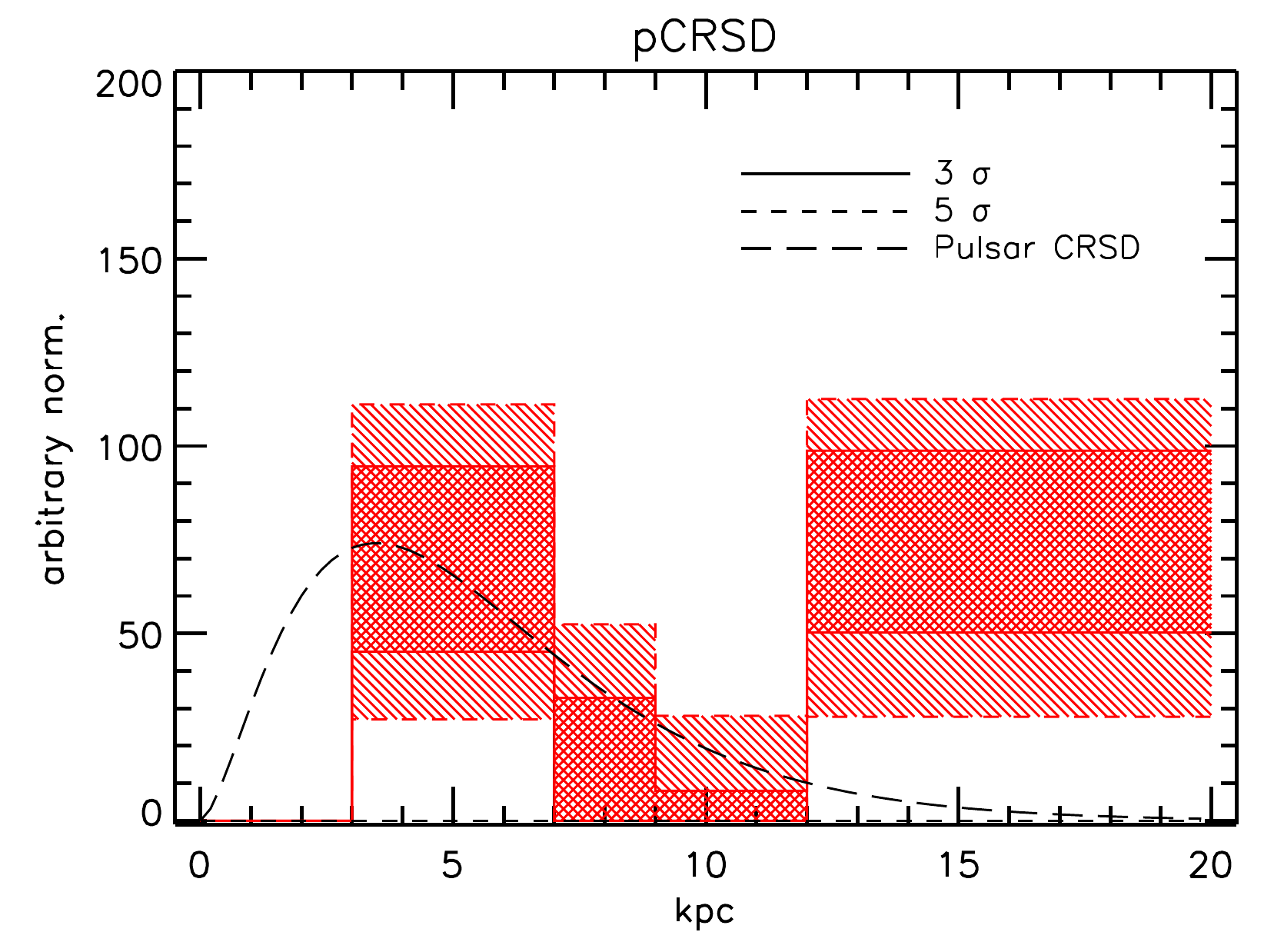}
\end{array}$
\end{center}
\caption{Best fit $e$ and $p$ CRSDs and errors, marginalized over the remaining parameters  and over the various DM models considered.  The pulsar distribution from \cite{2004A&A...422..545Y} is also shown for comparison. The source distribution is zeroed within 3 kpc of the Galactic center; see the text. \label{fig:CRSDs}}
\end{figure*}

In Figure \ref{fig:CRSDs} we show the CR source distributions derived in the fit (the coefficients $c_i^{e,p}$) and the uncertainty of them. Each bin is treated as a parameter for which the profile likelihood is built by marginalizing over all the other parameters (linear and nonlinear) and DM models and 3 and 5 $\sigma$ uncertainties are derived in the usual way. 
Again, since the Galactic Plane and the full energy range is not included in the fit, in interpreting the figure the same caveats as for Figure \ref{fig:profiles} apply. For example, the fitted $p$CRSD increases in the outer Galaxy: although protons from the outer Galaxy can propagate to the inner Galaxy (while few also propagate from the other side of the Galactic Center) and we thus have some sensitivity to them in our ROI, clearly, a proper statement on this feature would require to fit the outer Galaxy itself. 
Further caveats in interpreting these results include the fact that the CRSDs also depend on the ISM distribution and ISRF distribution, which are themselves affected  by large uncertainties. Finally, the CRSDs also depend on the chosen propagation setup. We do not attempt here to assess the systematic uncertainties related to these issues (see the next section, however, for some discussion regarding the ISRF uncertainties).

Overall, however, it can be noted that the $e$CRSD and $p$CRSD are in reasonable agreement with each other and with typical SNR or pulsars models, except for the rise in the outer Galaxy which is not predicted in these models. The plots are very similar when no DM contribution is assumed, which is expected since DM gives a subdominant contribution to the gamma-ray signal 3 kpc away from the GC.

An interesting point to comment on is how the DM constraints are affected if some of the nuisance parameters, in particular the CRSDs, are fixed to benchmark choices taken from the literature, instead of being marginalized away. We checked the resulting DM constraints for cases in which we fix the CRSDs (the $e$CRSD and $p$CRSD are taken to be equal) to three different common choices, namely SNR \citep{1998ApJ...504..761C}, pulsar \citep{2004A&A...422..545Y}, and a simple gaussian model (centered at the Galactic Center, with a width of 4.5 kpc) while the rest of the parameters is marginalized away in the same way as in the main analysis. We find that the SNR case gives slightly worse limits (20-30\%) while pulsar limits are very similar to the results of the main analysis (in agreement with the fact that the our best fit CRSDs are close to the pulsar distribution), and gaussian limits are a factor of $\leq$2 better than the pulsar case. This last case is understood in light of the fact that a gaussian CRSD,  being peaked at the Galactic Center, forces the inner Galaxy gamma-ray signal to be explained entirely in terms of ordinary CR sources, leaving little gamma-ray flux to DM and thus giving better limits.
This is interesting and might indicate that the DM constraints can become better if independent robust constraints on the CRSD become available.
On the other hand, as the results of the main analysis show, the fitted CRSD does not favor a Gaussian distribution (at least at a qualitative level; we have not performed a quantitative comparison for the reasons explained above).
For the time being, thus, the use of CRSD as a nuisance parameter seems the best approach, which, at the same time, leaves freedom in the fit to explore the degeneracy with DM and limit the fit to explore CRSDs which are in reasonable agreement with the (gamma-ray) data.

\begin{figure*}[tp] 
\begin{center}$
\begin{array}{cc}
%\hspace{-2pc}
\includegraphics[width=0.49\textwidth]{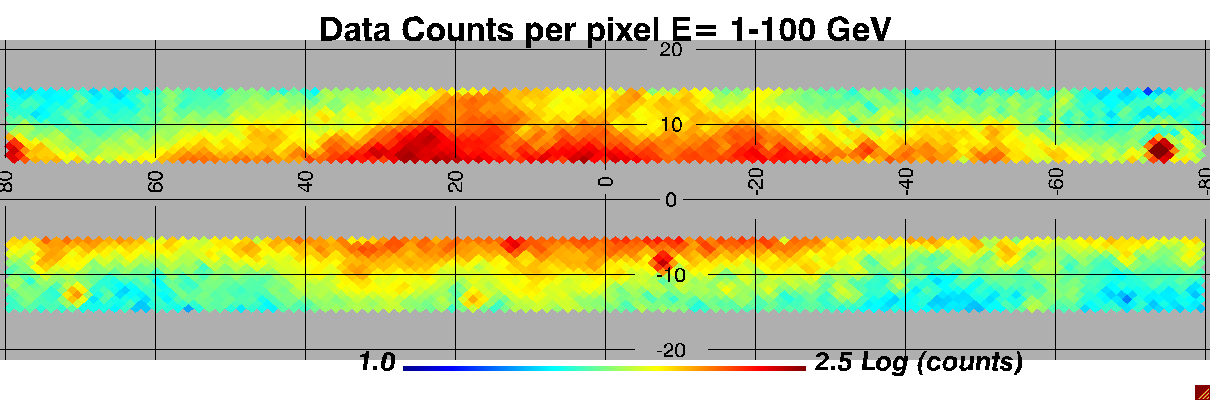}&
\includegraphics[width=0.49\textwidth]{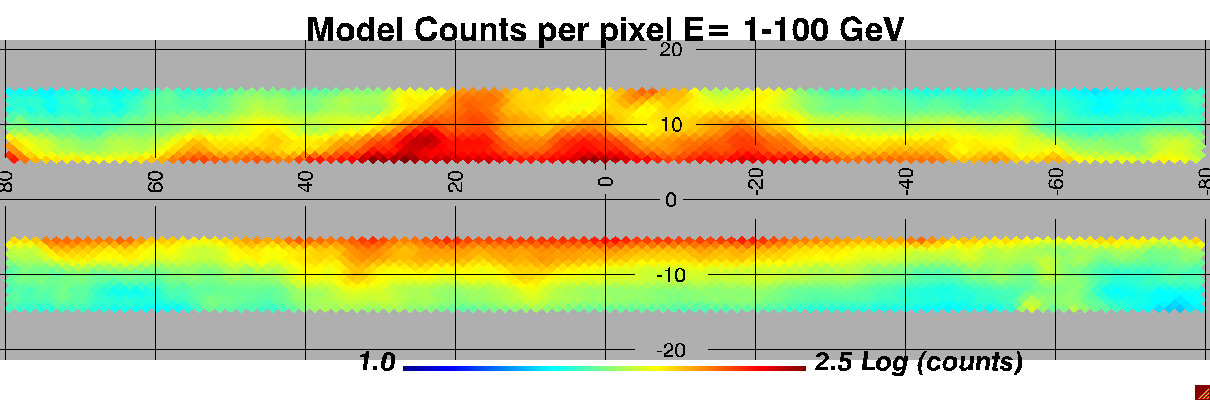}\\
\includegraphics[width=0.49\textwidth]{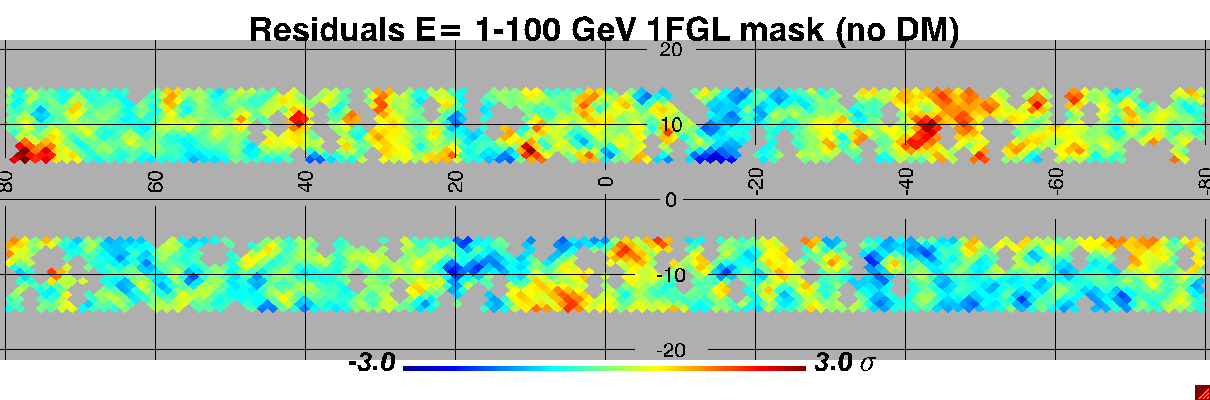}& 
\includegraphics[width=0.49\textwidth]{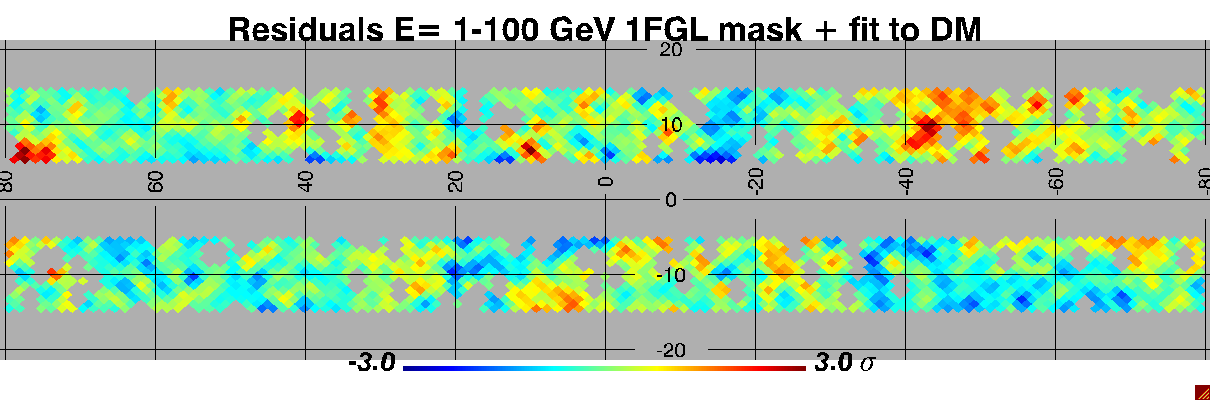}\\
\includegraphics[width=0.49\textwidth]{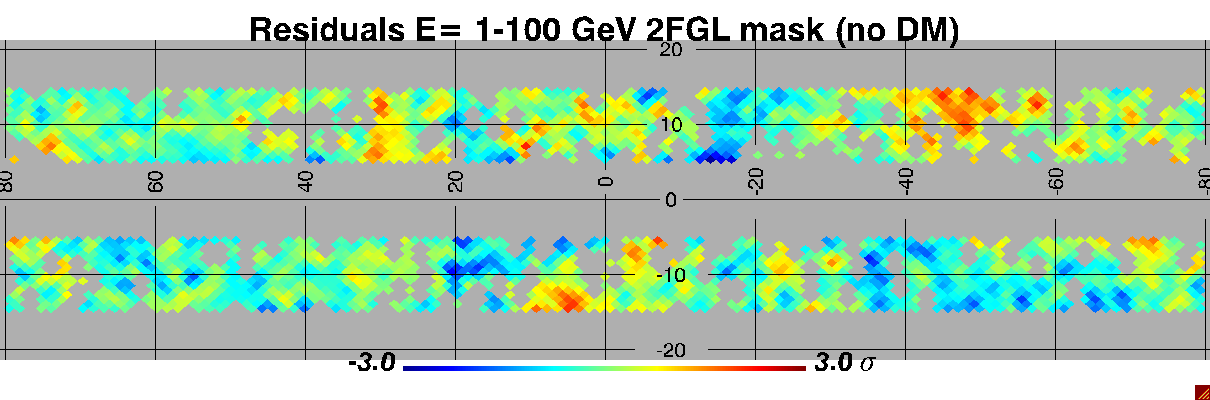}& 
\includegraphics[width=0.49\textwidth]{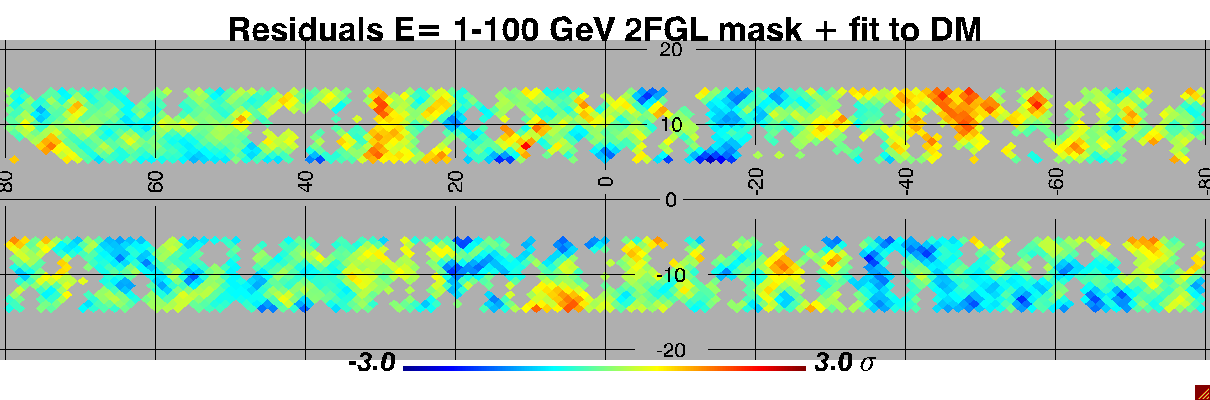}\\
\includegraphics[width=0.49\textwidth]{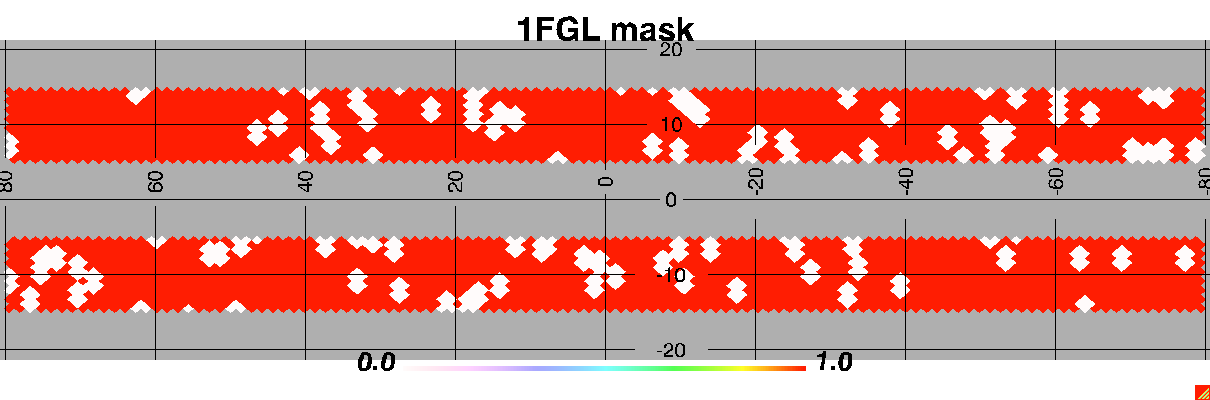}& 
\includegraphics[width=0.49\textwidth]{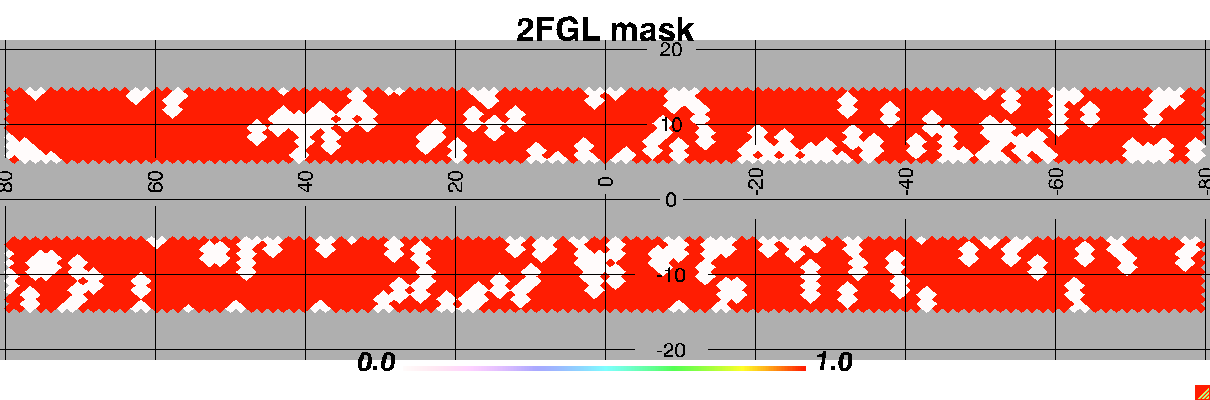}\\
\end{array}$
\end{center}
\caption{Counts map of the ROI that we consider (upper left panel), model prediction for a model (without DM) close to the best fit  ($z_h=10$ kpc, $\gamma_{e,2}=2.3$ and d2HI=0.0140 $\times 10^{-20}$ mag cm$^2$) parameter region (upper right), and residuals in units of $\sigma$ for the   same model (second row left) and when DM (of mass \mbox{$m_\chi$=150 GeV} and annihilating into $b\bar{b}$) is also included in the fit (second row right).  Third row: same as second row but with 2FGL point sources masked instead of 1FGL.  Fourth row: 1FGL mask (left) and 2FGL mask (right). The model and data counts and the residuals have been smoothed with a 1.25$^\circ$ Gaussian filter. The point sources mask in the residuals have been applied before and after the smoothing. \label{fig:residuals}}
\end{figure*}

Finally, we show in Figure \ref{fig:residuals} the counts map in our ROI, together with the model prediction for a model close to the best fit ($z_h=10$, kpc $\gamma_{e,2}=2.3$ and d2HI=0.0140 $\times 10^{-20}$ mag cm$^2$) and its residuals when DM (of mass \mbox{$m_\chi$=150 GeV} and annihilating into $b\bar{b}$) is included or not in the fit. It can be seen that the residuals are mostly flat, meaning that the model (and the models close to the minimum) are a reasonable fit to the data. A few features are however present, like the excess in the vicinity of ($l$,$b$)$\simeq$($-$45,10) and $\simeq$(7,$-$15) which seem to be related to the low latitude tip of Loop I and to the low latitude part of the South Lobe/Bubble, respectively. The two prominent negative residuals near ($l$,$b$)$\simeq$($-$15,5) and $\simeq$(20,$-$10), instead, approximately contour the Lobes and thus seem to be an artifact of the fit to compensate for this missing component. 
Gas misplaced in incorrect annuli also could be an alternative explanation.

We also show the point-source mask used based on the 1FGL catalog and, for comparison, the mask based on the 2FGL catalog \citep{2012ApJS..199...31N} and the residuals using this mask. Overall, it can be seen that the 2FGL mask covers few point sources which are apparent in the residuals with the 1FGL mask. The large scale features in the residuals are however unchanged, apart from a small part of Loop I near ($l$,$b$)$\simeq$($-$45,10) which is resolved into sources.

\section{Discussion on model uncertainties}\label{discussion}

In deriving our limits above we have taken into account many possible uncertainties like the ones in the $e,p$CRSDs, in $z_h$, the electron index and the dust to gas ratio.  We check below the importance of  further uncertainties which we have not considered explicitly in our scan.

An important component for which there is still a considerable uncertainty is the ISRF. In particular, the ISRF in the inner Galaxy is quite uncertain and the default model we used could be a substantial underestimate of the true one in this region. Very different ISRFs would affect the propagation of CREs through energy losses and this could be especially relevant for the DM models in which the IC component is important and provide strong constraints, like $\mu$ channels. Modes dominated by prompt radiation, like $b$ and $\tau$ should, instead, not be significantly subject to uncertainties in the ISRF. 
To make an explicit check we repeated our entire analysis using a different ISRF model \citep{PorterPrivate},  which has the bulge component  increased by a factor of 10 (see \cite{2008ApJ...682..400P} for a detailed definition), which implies an overall increase in the inner Galaxy of a factor of 2. The DM limits with this enhanced ISRF were, however, not appreciably affected.  We verified that the enhanced ISRF produces an enhanced IC component, but only within a few degrees of the Galactic Center, thus not affecting the fit in our ROI.  It also should be stressed that a more intense ISRF  implies more IC emission for the DM IC too, so that assuming a lower ISRF gives conservative limits. Finally, an ISRF lower than the one assumed here is also possible, as the results  obtained in \cite{2012arXiv1202.4039T} (see Figure 11 there) for the CRSD following the pulsar distribution seems to indicate. However, the ``ISRF normalization'' reported in \cite{2012arXiv1202.4039T}, is more precisely a proxy for a combination of ISRF intensity, normalization of the CRE spectrum and halo size, so that alternative explanations are possible. 

We also checked more systematically other sources of uncertainties, but in a more simplified setup:   we set a particular model as  reference and then we varied each parameter one at a time, keeping the others fixed,  and for each case we calculate the percentage variation in DM limits for selected DM models. 
We vary the parameters derived from the CR fit,  $v_A$, $\gamma_{p,1}$, $\gamma_{p,2}$,  $\rho_{br,p}$, and the $(D_0,z_h)$ relation, within the uncertainty ranges derived in \cite{2012arXiv1202.4039T} enlarging it by a factor of $\sim$2 to take into account possible systematic uncertainties (the errors quoted in  \cite{2012arXiv1202.4039T} are statistical only).  
We also include in the list of the tested parameters the ones which are included in our model scan (CRSD, d2HI, $\gamma_{e,2}$, $(D_0,z_h)$)  to allow for a direct comparison.
The following set of parameters, which lie close to the best fit of our analysis, was chosen for the reference model: $v_A=$36 km s$^{-1}$,  $D_{0}=5.0~10^{28}$cm$^2 $s$^{-1}$, $z_h=4$ kpc, $\delta=0.3$,  $\gamma_{p,1}=1.9$,  $\gamma_{p,2}=2.39$,  $\rho_{br,p}=11.5$, $\gamma_{e,2}=2.45$, d2HI=0.014 $\times 10^{-20}$ mag cm$^2$, CRSD=SNR, $V_c$=0 km s$^{-1}$.  Results are shown in Table \ref{tab:smallgrid}.

We can see that CR parameters such as $v_A$ and $\gamma_{p,1}$, $\gamma_{p,2}$,  $\rho_{br,p}$, $V_c$ and even  different gas maps have very low ($\lsi 10\%$) impact on the DM limits.
The table confirms that $\gamma_{e,2}$ and the CRSD (which we fix here to be the same for protons and CREs) are the main parameters degenerate with DM and thus affecting the limits the most (up to $60\%$). 
The diffusion constant $D_0$ is tightly correlated to the halo height $z_h$. Therefore we
vary the parameter pair ($D_0$, $z_h$) instead of the single parameters, using their relation derived from the fit to the CR data described in Sec. \ref{diffusemodeling}. Nonetheless, the combination $D_0$ and $z_h$ are included individually in the parameter scans of the previous section used for the main results.  
As an additional check of the effect of the CR propagation parameters on the DM limits, we find DM limits in three theoretical CR propagation setups: plain diffusion (PD, characterized by index of diffusion of $\delta =0.6$), Kraichanian (KRA, $\delta =0.5$), and Kolmogorov (KOL, $\delta =0.3$). In these cases, the rest of the CR propagation parameters are found from the best fit to the CR data following the method described in \cite{2012arXiv1202.4039T,2011APh....34..528D}. These fits to CR are performed again without any DM component. We find that DM limits in these three CR diffusion setups are also barely affected, in particular when compared to the effect of the CR source distribution, as shown in Table \ref{tab:smallgrid}. 

Finally we also consider an alternative configuration of the Galactic Magnetic Field (GMF). The reference one is the default configuration used in \texttt{GALPROP} with an exponential profile in $R$ and $z$ and length scales of 10 kpc in $R$ and 2 kpc in $z$, normalized to 5 $\mu$G locally (Conf~1). The alternative configuration we tested has in addition a further component of constant 100 $\mu$G intensity within 0.4 kpc from the Galactic Center, as motivated by a recent work \cite{2010Natur.463...65C} (Conf~2). This alternative configuration also produces changes in the limits of less than 10\%.

\begin{table*}[t]
\begin{center}
\begin{tabular}{ | c|c|c| }
\hline
 \textbf{ {Parameter}} & \textbf{  {$|\delta \sigma /\sigma|$ [$\%$]}}, $b\bar{b}$  & \textbf{  {$|\delta \sigma /\sigma|$ [$\%$]}}, $\mu^+\mu^-$ \\
\hline \hline
 {\bf $v_A$} [~30;~{\bf 36};~45] km~s$^{-1}$& [~6;~{\bf 0};~11]      & [~4.;~{\bf 0};~9] \\
 {\bf $\gamma_{p,1}$} [~1.8;~{\bf 1.9};~2;] & [~1.0;~{\bf 0};~2.5]   & [1.5;~{\bf 0};~2.0]\\
 {\bf $\gamma_{p,2}$} [~2.35;~{\bf 2.39};~2.45] & [~2.5;~{\bf 0};~1.5]   & [2.5;~{\bf 0};~1.5]\\
  {\bf $\rho_{br,p}$} [~10;~{\bf 11.5};~12.5] GV & [~0.5;~{\bf 0};~1.0]   & [0.9;~{\bf 0};~1.5]\\
{\,\, d2HI} [~0.0110,~{\bf  0.0140};~0.0170] $10^{-20}$ mag cm$^2$ \,\, & [3;~{\bf 0};~12]   &  [~3;{\bf 0};~9] \\
\hline
\hline
 
 {\bf $\gamma_{e,2}$} [~2.0;~{\bf 2.45};~2.6]& [~17;~{\bf 0};~7]         &  [~18;~{\bf 0};~5]   \\
 {\bf $(D_0,z_h)$} [~{\bf (5.0e28, 4)};~{ (7.1e28, 10)}] cm$^2$s$^{-1}$ & [~{\bf 0};~10]    &  [~{\bf 0};~7] \\
 {CRSD} [~{\bf SNR};~Pulsar] & [~{\bf 0};~61]   &  [~{\bf 0};~59]\ \\
\hline
\hline

 ~KRA($\delta=0.5$);~{\bf KOL($\delta=0.3$)};~PD($\delta=0.6$) & [~4.0;~{\bf 0};~3.0]   & [1.0;~{\bf 0};~5]\\
 {\bf $V_c$} [{\bf 0};~20] km~s$^{-1}$& [~{\bf 0};~6]  &  [~{\bf 0};~4] \\
 { GMF} [~{\bf Conf~1},~Conf~2] &  [~{\bf 0};~3]&     [~{\bf 0};~8]\\

 \hline
\end{tabular}
\caption{Relative variation $|\delta\sigma/\sigma|$[$\%$]  of the limits on the DM velocity averaged annihilation cross-section derived in this work with respect to changes in the underlying astrophysical diffuse emission model. The table shows the relative variation for selected DM models ($b{\bar b}$ and $\mu ^+ \mu ^-$ channel, for a 150 GeV DM)  in a simplified set-up  when only one parameter is varied at a time. Each row corresponds to the indicated parameter. The bold values correspond to the reference value.}
\label{tab:smallgrid}
\end{center}
\end{table*}

\section{Summary} \label{summary}

In this work we constrain the contribution to the diffuse gamma-ray emission from DM annihilating or decaying in the Milky Way halo, based on the {\it Fermi} LAT gamma-ray data. We first present the most conservative limits on DM assuming that all LAT photons from the Halo are produced by annihilating/decaying WIMPs. Then, based on our current best knowledge of the Galactic diffuse emission \citep{2012arXiv1202.4039T}, we use \texttt{GALPROP} to model the astrophysical diffuse background, and, using a profile likelihood approach, we explore the effects of various poorly constrained parameters in
the modeling of the astrophysical background, e.g., the diffusive halo height, the CR source distribution, the index of the electron injection spectrum and the dust to gas ratio in order to get more robust constraints.
We also remove astrophysical CR sources within 3 kpc  from the Galactic Center so that any potential astrophysical contribution in this region is attributed to DM, resulting in more conservative constraints.
Overall, rather than being due to residual astrophysical model  uncertainties, the remaining major uncertainties in the DM constraints from the Halo region come from the modeling of the DM signal itself.
The main uncertainty is in the normalization of the DM profile, which is fixed through the local value of the DM density. We use the recent determination $\rho_0=0.43$ GeV cm$^{-3}$ from \cite{2010A&A...523A..83S}, which has, however, a large uncertainty, with values in the range 0.2-0.7 GeV cm$^{-3}$ still viable. A large uncertainty in $\rho_0$ is particular important for annihilation constraints since they scale like $\rho_0^2$, while for constraints on decaying DM the scaling is only linear.
A less important role is played by the uncertainties in the DM profile, since in the Halo region different profiles predict similar DM densities.
When using the lowest allowed DM density used in literature of $\rho_0=0.2$ GeV cm$^{-3}$ (see e.g. \citealp{2010A&A...523A..83S}), our limits worsen by a factor 4 (2) for annihilating (decaying) DM, as shown in Figures \ref{fig:fixedsourcelimits} and \ref{fig:decay}. A better determination of the local DM density, as well as of the parameters determining the global structure of the DM Halo, is therefore of the utmost importance for reducing the uncertainties related to DM constraints from DM halo, but it is beyond the scope of this paper and is the subject of dedicated studies.

Bearing this in mind, the limits we obtain are competitive with complementary probes of DM like dwarfs, clusters of galaxies, or recent constraints obtained from CMB observations for DM models with prompt spectra, and significantly improve over these studies for DM models with significant IC contribution such as DM annihilating into $\mu^+\mu^-$.  The limits we derive for leptonic models  challenge the interpretation of the PAMELA and {\it Fermi} CR anomalies as annihilation of DM in the Galactic Halo, while they are not constraining enough to exclude the interpretation in terms  of decaying DM. We note that this last conclusion is not affected by the uncertainty in $\rho_0$ since both the derived constraints and the region of parameter space compatible with the DM interpretation of the CR anomalies scale in the same way with $\rho_0$.

An obvious improvement of this analysis would be a full scan of the CR parameters from a simultaneous fit to both gamma and cosmic-ray data. An effort in this direction is currently ongoing and will be  reported in future publications.

\section*{Acknowledgements}

The \textit{Fermi} LAT Collaboration acknowledges generous ongoing support from a number of agencies and institutes that have supported both the development and the operation of the LAT as well as scientific data analysis. These include the National Aeronautics and Space Administration and the Department of Energy in the United States, the Commissariat \`a l'Energie Atomique and the Centre National de la Recherche Scientifique / Institut National de Physique Nucl\'eaire et de Physique des Particules in France, the Agenzia Spaziale Italiana and the Istituto Nazionale di Fisica Nucleare in Italy, the Ministry of Education, Culture, Sports, Science and Technology (MEXT), High Energy Accelerator Research Organization (KEK) and Japan Aerospace Exploration Agency (JAXA) in Japan, and the K.~A.~Wallenberg Foundation, the Swedish Research Council and the Swedish National Space Board in Sweden.
\\ \indent
Additional support for science analysis during the operations phase is gratefully acknowledged from the Istituto Nazionale di Astrofisica in Italy and the and the Centre National d'\'Etudes Spatiales in France.
\\ \indent
We acknowledge useful discussions with Illias Cholis, William Guillard, Pasquale Serpico, Paolo Salucci, Joakim, Edsj\"{o} and are grateful to Marco Cirelli for providing a plotting script for the PAMELA/{\it Fermi} regions, used in figures \ref{fig:fixedsourcelimits} and \ref{fig:decay}. Some of the results in this paper have been derived using the HEALPix \citep{2005ApJ...622..759G} package.

\clearpage

\bibliographystyle{apj}
\bibliography{resubmitted_apj}

\end{document}